\documentclass[12pt,draftclsnofoot,onecolumn]{IEEEtran}
\usepackage{hyperref}
\usepackage{color}
\usepackage{amsmath}
\usepackage{bm}
\usepackage{mathrsfs}
\usepackage{multirow}
\usepackage{subfig}
\usepackage{graphicx}
\usepackage{amssymb}
\usepackage{calc}
\usepackage{stfloats}
\usepackage{comment}
\usepackage{enumerate}
\graphicspath{{figures/}}
\usepackage{algorithm}
\usepackage{algorithmic}
\usepackage{textcomp}
\usepackage{gensymb}

\makeatletter
\def\endthebibliography{%
	\def\@noitemerr{\@latex@warning{Empty `thebibliography' environment}}%
	\endlist
}

\begin{document}

\title{\huge Measurement-Level Fusion for OTHR Network Using Message Passing}
	
\author{Hua Lan\thanks{Hua Lan, Zengfu Wang, Xianglong Bai, Quan Pan are with the School of Automation, Northwestern Polytechnical University, and the Key Laboratory of Information Fusion Technology, Ministry of Education, Xi'an, Shaanxi, 710072, PR China. Kun Lu is with Nanjing Reasearch Insitute of Electronics Technology, Nanjing, Jiangsu, 210039, China. Zengfu Wang is also with Faculty of Electrical Engineering, Mathematics and Computer Science, Delft University of Technology, Delft 2826 CD, the Netherlands.}, Zengfu Wang, Xianglong Bai, Quan Pan, Kun Lu}

\maketitle

\vspace{-0.5cm}

\begin{abstract}
Tracking an unknown number of targets based on multipath measurements provided by an over-the-horizon radar~(OTHR) network with a statistical ionospheric model is complicated, which requires solving four subproblems: target detection, target tracking, multipath data association and ionospheric height identification.
A joint solution is desired since the four subproblems are highly correlated, but suffering from the intractable inference problem of high-dimensional latent variables.
In this paper, a unified message passing approach, combining belief propagation~(BP) and mean-field~(MF) approximation, is developed for simplifying the intractable inference.
Based upon the factor graph corresponding to a factorization of the joint probability distribution function~(PDF) of the latent variables and a choice for a separation of this factorization into BP region and MF region, the posterior PDFs of continuous latent variables including target kinematic state, target visibility state, and ionospheric height, are approximated by MF due to its simple MP update rules for conjugate-exponential models.
With regard to discrete multipath data association which contains one-to-one frame~(hard) constraints, its PDF is approximated by loopy BP.
Finally, the approximated posterior PDFs are updated iteratively in a closed-loop manner, which is effective for dealing with the coupling issue among target detection, target tracking, multipath data association, and ionospheric height identification.
Meanwhile, the proposed approach has the measurement-level fusion architecture due to the direct processing of the raw multipath measurements from an OTHR network, which is benefit to improving target tracking performance.
Its performance is demonstrated on a simulated OTHR network multitarget tracking scenario.
\end{abstract}



\section{Introduction}
By exploiting sky-wave propagation via reflection by the ionosphere, an over-the-horizon radar~(OTHR) provides detection and monitoring of both air targets and maritime targets beyond the line-of-sight horizon~\cite{Fabrizio2013}.
Due to its cost-effectiveness and the ability to monitor remote geographical regions where conventional line-of-sight radars are not able to cover, OTHR has been received long-standing interest in both defense and civil applications.
An OTHR network, such as Australia's Jindalee Operational Radar Network~(JORN)~\cite{anderson1986remote},
consists of multiple OTHRs that have a degree of overlapping coverage and are operated jointly to achieve overall mission objectives.
Compared with a single OTHR, many benefits can be achieved through fusing all the information from an OTHR network:
(1)~\emph{Improved target detection:}
Target detection ability of an OTHR is related to the fading characteristics of ionosphere and the Doppler speed of a target.
Multiple and independent OTHRs have different ionospheric fading patterns and geometric positions relating to the target.
Therefore, integrating the information from an OTHR network can improve detection probability of the target, and this also benefits to increasing the timeliness of initializing tracks.
(2)~\emph{Improved target tracking:}
Once target detection probability is improved, target track detection probability and stability of target tracking will be improved as well. Moreover, redundant data from an OTHR network increases target track accuracy by providing independent observations of the target.
(3)~\emph{Better coordinate registration~(CR):} Improved target track accuracy can aid in resolving ambiguities and uncertainties in the decision of multipath data association~\footnote{There are two types of association in OTHR, i.e., data association due to multiple targets scenario and path association arising from multipath propagation. We call the target-to-measurement-to-path association as multipath data association.} and the identification of ionospheric height.

However, because of the ionosphere, which is complex in nature,
difficulties arise from both continuous and discrete uncertainties for an OTHR network fusion:
(1) \emph{CR uncertainty:} CR uses available information on ionosphere, including propagation paths and corresponding ionospheric heights, to localize a target by converting observations of the target in radar slant coordinate systems to ground or geographic coordinate systems.
However, high-frequency signal propagation through separate refractive layers in the ionosphere often results in multiple propagation paths/modes between a target and an OTHR, producing multiple resolved~(multipath) measurements of the target with a high probability.
Accordingly, path/mode association ambiguity occurs in CR.
Moreover, ionospheric heights vary spatially and temporally, adding an extra degree of uncertainty to CR.
If the propagation path is not selected correctly and/or ionospheric heights are not estimated precisely,
the ground track of a target will be inaccurate and it will be difficult to correlate the ground tracks from multiple OTHRs.
(2) \emph{Data association uncertainty:} The OTHR performance characteristics are indicated by poor measurement accuracy, long sampling period, low detection probability~(per path), and high false-alarm rate, complicating the data association.
In a single OTHR tracker, ghost tracks will arise if multipath measurements are not be associated correctly with the underlying target in the tracking stage, or multipath tracks are not be fused correctly at the post-tracking stage.
This problem is exacerbated when an unknown number of targets are tracked in an OTHR network.

Achieving multitarget tracking by fusing unlabeled multipath measurements from an OTHR network,
requires solving four subproblems: target detection, target tracking, multipath data association,
and ionospheric height identification.
Most existing target tracking algorithms are only applicable to a single OTHR.
The multipath track fusion algorithm~(MPTF)~\cite{Percival1998} reflects the view that tracking and fusion are two-stage process whereas the first-stage tracking process produces the multipath slant tracks independently,
and the second-stage fusion process associates those slant tracks and fuse them.
Multipath measurement fusion approaches~\cite{pulford2004othr, pulford1998multipath, Habtemariam2013, sathyan2013multiple, Xu2015, Chen2014, Lan2014TSP, Lan2014Fusion} integrate tracking and fusion as a single, unified process.
By extending the existing data association approach to multipath data association, the target tracks in ground coordinate systems are updated directly using multipath measurements.

Extending MPTF for a single OTHR to multiple OTHRs~(MR-MPTF), the work of~\cite{rutten2003track, sarunic2001over} studied fusion for an OTHR network.
Based on (slant)~track-level fusion framework, MR-MPTF treats target tracking and multiple radar fusion as two independent processes.
Specifically, MR-MPTF carries out single-path target tracking in radar slant coordinate system for each OTHR,
generates multiple feasible association hypotheses using all possible combinations of existing ionospheric paths,
transform the relevant multipath tracks from multiple OTHRs into a common coordinate system whereas all track-to-target association hypotheses are recursively constructed, and then fuse target states with the weight given by the probability of each hypothesis.
MR-MPTF has the practical benefits of incrementally augmenting an OTHR by adding a second independent fusion function on top of the existing tracking modules.
However, the fusion performance of MR-MPTF heavily relies on the tracking performance of each single OTHR.
If the detection performance of an OTHR is poor, the multipath tracks provided by the OTHR will be inaccurate and intermittent, or even missed.
In this case, MR-MPTF may encounter problems of unreliable track fusion and/or ghost tracks.
Moreover, MR-MPTF might be time-consuming due to the multiple hypothesis nature~\cite{rutten2003track}.

Different from the above-mentioned track-level fusion framework,
measurement-level fusion framework performs target detection, tracking, and fusion on the raw measurements of sensors.
Comparing with the track-level fusion framework, the measurement-level fusion framework reduces information loss and has the advantages of improving the performance of target tracking, especially in low signal-to-noise ratio~(SNR) environments.
In principle, the problem of measurement-level fusion for an OTHR network can be formulated as an intractable inference problem.
This is because the required probabilistic models involve both discrete and continuous latent variables with high-dimension, such as target visibility state, target kinematic state, multipath data association, and ionospheric height. In such case, one needs to resort to approximation methods.
Mean-filed~(MF) approximation~\cite{Blei2016variational} and (loopy) belief propagation~(BP)~\cite{yedidia2003understanding} are two kinds of approximation methods that scale well to high-dimensional inference problems.
BP computes the marginal distribution of a certain joint probability distribution function~(PDF) by minimizing Bethe free energy,
while MF approximates a joint PDF through minimizing variational free energy.
Both methods have their pros and cons \cite{riegler2012merging}.
By the factorization assumption where the latent variables are mutually independent, MF yields closed-form tractable expressions and admits a convergent implementation, in particular for conjugate-exponential models. However, being limited by its strong factorization,
MF cannot capture the dependencies between latent variables,
which makes it incompatible with hard constraints where the dependencies are of intrinsic interest.
BP is compatible with hard constraints, but when being applied to probabilistic models that involve both discrete and continuous latent variables, it may have high complexity.
Both MF and BP can be implemented by message passing~(MP) on the factor graph, i.e., variable nodes pass messages to factor nodes and factor nodes pass messages to variable nodes. This iterative process is repeated until the messages converge to a fixed point.
To exploit their respective virtues and circumvent their drawbacks, BP and MF are combined in a unified MP algorithm~\cite{riegler2012merging} based on region-based free energy approximations~\cite{yedidia2005constructing}.
Recently, MP has been attracting much attention from the target tracking community benefiting by its estimation accuracy, computational efficiency and implementation flexibility~\cite{meyer2018message}.
For the first time, Turner \emph{et al.}~\cite{Turner2014} proposed a complete variational tracker that integrates the target detection, target tracking and data association in a unified Bayesian framework, and the intractable Bayesian inference is approximated by MF.
Inspired by the work of~\cite{Turner2014}, Lan~\emph{et al.}~\cite{Lan2015IRS} proposed an MF-based multipath multitarget tracking algorithm that integrates multipath measurements to improve the performance of both detection and tracking.
Williams and Lau~\cite{Williams2014} addressed the data association problem based on BP.
By formulating data association problem as an inference problem on the graphical model, the marginal association probabilities are approximated by loopy BP, and the convergence of BP for data association problem was proved.
The multiple scan data association problem was considered in~\cite{Willams2016},
where a convex free energy was constructed and optimized by a primal-dual coordinate ascent method.
The multisensor-multitarget tracking problems were considered in~\cite{meyer2017scalable}, where the marginal PDFs of target detection, target tracking, and data association are approximated by running particle-based implementation of BP on a suitably devised factor graph.
The extension of~\cite{meyer2017scalable} to time-varying parameters, such as detection probability and multiple dynamic models, was proposed in~\cite{soldi2019self}. However, the existing MP-based multitarget tracking algorithms are either from the view of variational optimization~(MF approximation)~\cite{Turner2014,Lan2015IRS} or from the view of BP methods~\cite{Williams2014, Willams2016, meyer2017scalable, soldi2019self}.
None of them is based on a unified MP method that integrates both MF and BP.

This paper present an MP-based measurement-level fusion approach for an OTHR network, referred as MP-OTHRs.
MP-OTHRs carries out the intractable inference of joint high-dimensional latent variables, including target visibility state, target kinematic state, multipath data association and ionospheric height in a unified optimization procedure.
The interdependence of the high-dimensional latent variables is modeled by a factor graph, which is divided into an MF-part and a BP-part.
Accordingly, the MF-part that contains the conjugate-exponential latent variables including target visibility state, target kinematic state, ionospheric height performs the message passing using the MF update rules, and the BP-part including the latent variables of multipath data association which fulfills one-to-one hard constraints performs the message passing using the BP update rules.
The beliefs, i.e., approximated posterior PDFs, are updated iteratively by message passing on the factor graph. This iterative process is repeated until all the beliefs converge to a fixed point. Finally, the problems of target detection, target tracking, multipath data association and ionospheric height identification are solved in a unified MP framework. This joint solution is especially important because the output of each of the problems are strongly correlated and the solution of one can help improve the others.
Meanwhile, MP-OTHRs integrates raw multipath measurements from all OTHRs to enhance the performance of both target detection and tracking, especially for weak targets. In summary, our novelties and contributions are as follows:
\begin{itemize}
\item For the first time, we develop a measurement-level fusion approach for multiple target tracking for an OTHR network with a statistical ionospheric model.
\item We provide a unified MP approach that combines MF and BP approximation to solve the problems of target detection, target tracking, multipath data association and ionospheric height identification simultaneously.
\end{itemize}

The remainder of the paper is organized as follows.
Section~\ref{sec:PF} describes the problem formulation of an OTHR network measurement-level fusion.
Section~\ref{sec:solutions} introduces the proposed MP-OTHRs algorithm based on the combined MF-BP method.
In Section~\ref{sec:simulation}, the simulation comparison with the track-based MR-MPTF algorithm is given.
Finally, Section~\ref{sec:conclusion} concludes the paper.
	
\section{Problem Formulation}\label{sec:PF}
\begin{figure}[!htbp]
\centering
\includegraphics[scale = 0.5]{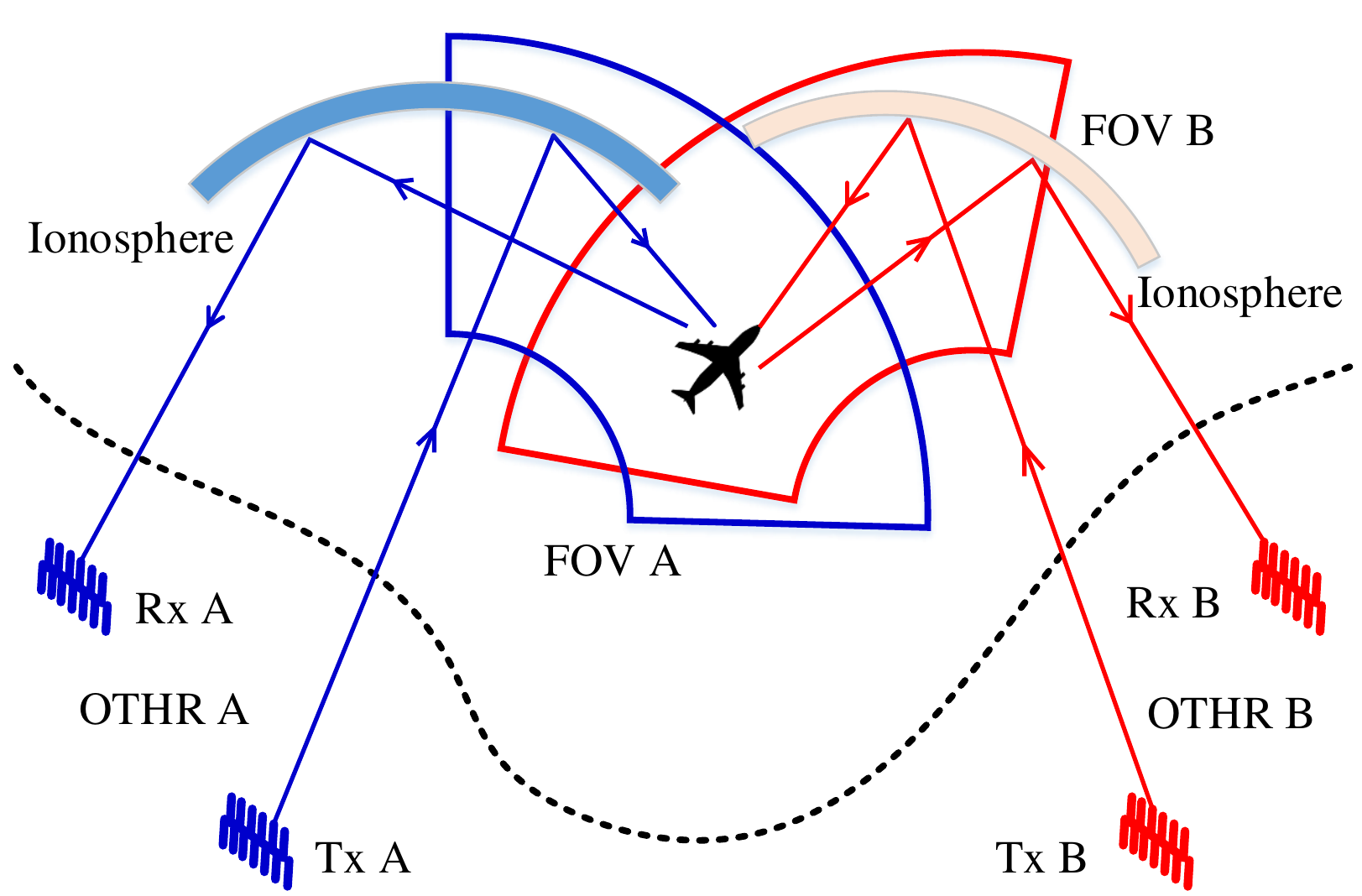}
\caption{Illustration of the operational model for an OTHR network consisting of two OTHRs~\cite{zhang2007concurrent}.
OTHR-A and OTHR-B, located at different sites, provide independent observations of a partially overlapping region to detect and track targets.}
\label{1-Scenario}
\end{figure}

This paper addresses the problem of joint target detection and tracking using multipath measurements provided by an OTHR network.
Figure.~\ref{1-Scenario} illustrates the operational model of an OTHR network consisting of two OTHRs.
We assume that every OTHR in the network sends its measurements to a fusion center,
where all the measurements are processed.
In this section, we first describe the statistical models of the target, the sensor measurement, the ionospheric environment,
and then introduce multipath data association.
At last, we state the problem of the measurement-level fusion to be solved for an OTHR network in the Bayesian framework.

\subsection{Target Kinematic State and Visibility State Modeling}
MR-MPTF~\cite{sarunic2001over} adopts a centralised track-level fusion framework.
Specifically, MR-MPTF models target kinematic state and produces multipath tracks in the local (and noninertial) slant coordinate system of each OTHR, and then transforms all the local multipath tracks from all OTHRs into a common coordinate system, e.g., the ground coordinate system of an OTHR. Lastly, the single radar MPTF algorithm is applied.
Distinct from MR-MPTF, we model target kinematic state in an inertial coordinate system--universal transverse Mercator~(UTM) coordinate system~\cite{grafarend1995optimal}.
The reason is as follows.
Due to the low-range resolution associated with the narrowband radar signal and inaccuracy in the estimated ionosphere parameters,
the direct estimation of target altitude is difficult and imprecise for OTHR~\cite{zhang2019target}.
As a result, the target kinematic state is common modeled in a plane coordinate system.
The targets of interest move in inertial space following straight lines with small deviation from this model~\cite{bar1990multitarget}.
This being the case, the most natural choice of a coordinate system is a Cartesian system that allows targets to be modeled as a linear state equation.
The UTM coordinate system, which can be regarded as a two-dimensional Cartesian system~(X-Y plane), is a horizontal position representation of the geodetic inertial coordinate system~(latitude-longitude), ignoring altitude information.
Compared with other coordinate systems that commonly used in OTHR target tracking, such as geodetic coordinate system and ground coordinate system, it is more accurate for modeling the target kinematic state with a linear state equation in a UTM coordinate system.

The joint kinematic states of all targets at time $k$ are denoted by $X_k = \big\{x_{i,k}\big\}_{i = 1, \cdots, n_k^x}$,
where $n_k^x$ is the maximum possible number of targets in the overlapping region of an OTHR network.
The kinematic state of each target at time $k$ is represented in UTM coordinate system $\mathcal{MX}$ as $x_{i,k} = [x_k, \dot x_k, y_k, \dot y_k]^T$, $i = 1, \ldots, n_k^x$, which consists of the $i$th target's position $x$, $y$ and velocity $\dot x$, $\dot y$.
Each target evolves independently according to the following linear dynamical equation
\begin{equation}\label{1-State-X}
x_{i, k+1} = F_k x_{i, k} + v_k, \quad i = 1, \cdots, n_{k}^x,
\end{equation}
where $F_k$ is the state transition matrix, $v_k$ is the zero-mean white Gaussian process noise with known covariance matrix $Q_k$.
The initial kinematic state of each target $x_{i,0}$, in general unknown, is modeled as a Gaussian-distributed random vector.

The joint visibility~(detection) state of all targets at time $k$ are denoted by $E_k = \{e_{i,k}\}_{i = 1, \cdots, n_k^x}$ with the binary variable $e_{i,k} \in \{0, 1\}$ representing the visibility state of target $i$ at time $k$.
In the vein of \cite{pulford1998multipath}, target $i$ is visible~(detectable) at time $k$ if $e_{i,k} = 1$,
and is invisible~(undetectable) if $e_{i,k} = 0$.
The evolution of the visibility state $e_{i,k}$ is modeled as a two-state Markov process
\begin{equation}\label{2-State-E}
\text{Pr}(e_{i, k+1}) = T_{k}\text{Pr}(e_{i, k}), \quad i = 1, \cdots, n_k^x,
\end{equation}
where $T_{k} = \text{Pr}(e_{i, k+1}|e_{i, k})$ is the known transition probability, and the initial probability $\pi_i = \left[\text{Pr}(e_{i,0} = 0), \text{Pr}(e_{i,0} = 1)\right]^T$.

\subsection{Ionospheric Height Modeling}
The ionosphere, the medium that the high-frequency radar signals are propagated in, is a broad layer of ionized gas located from 60 to 1000~km above the earth's surface. It can be divided into several subregions including D region~(50-90~km), E region~(90-140~km), and F region~(140-400~km).
The ionization in D layer is very low such that it does not contribute to the reflection of OTHR signals.
Let $\hbar_t$ and $\hbar_r$ be the ionospheric heights where the transmitting and the receiving signals are reflected by, respectively.
Each pair of ionospheric heights $u_{\tau} = \left[\hbar_{t, \tau}, \hbar_{r, \tau}\right]^T, \tau = 1, ..., n^m$ represents a specific propagation path, where $n^m$ is the known number of possible propagation paths.
Multipath propagation phenomenon gives rise to multiple resolved target-originated measurements independently.

These ionospheric parameters are typically derived through the ray-tracing technique by combining with an empirical ionospheric model, where the model parameters including the vertical electron density or plasma frequency profile are measured or estimated by the ionosondes subsystems consisting of a network of quasi-vertical and wide-sweep backscatter ionograms.
Note that the ionospheric parameters, which are radar-specific, are related to the geographic location and operation frequency of an OTHR.
Due to the long interval between soundings, the limited spatial resolution for typical wide-sweep backscatter ionograms and the empirical modeling error, the ionosondes provide spatially and temporally incomplete information about the ionosphere~\cite{anderson2002track}.
As a result, errors in the identification of ionospheric parameters, including the propagation path $\tau$ and the ionospheric height $u_{\tau}$, can seriously degrade the performance of target tracking.

To model the temporal uncertainty of ionospheric height, a statistical ionospheric model is considered by assuming that the state of ionospheric height $u \in \mathcal{IH}$ is Gaussian-distributed as
\begin{equation}
u_{k+1} = B_{k}u_{k} + q_k,
\end{equation}
where $B_k$ is the known state transition matrix of ionospheric height,
$q_k$ is the corresponding zero-mean white Gaussian process noise with covariance matrix $\mathcal{Q}_k$.

The corresponding measurement equation is given as
\begin{align}
I_{k+1} = C_{k+1}u_{k+1} + \nu_{k+1},
\end{align}
where $C_k$ is the ionospheric measurement matrix, and $\nu_k$ is a zero-mean Gaussian white noise with covariance $W_k$.

\subsection{OTHR Measurement Modeling}
Assume that there are $n^s$ OTHRs in the OTHR network. Each OTHR $s$ consists of the primary radar system to detect the targets of interest in the radar slant coordinate system, and the secondary ionosondes to measure the ionospheric height.
The measurements of all OTHRs at time $k$ is denoted as $Y_k = \{Y_k^s\}_{s = 1, \cdots, n^s}$ with $Y_k^s = \big\{y_{j, k}^s\big\}_{j = 1 \cdots n_k^{e,s}}$ being the measurement set of the $s$th OTHR, where $n_k^{e,s}$ is the number of measurements.
For the $s$th OTHR, each measurement in radar slant coordinate system $\mathcal{RS}$ at time $k$ is represented as $y_{j,k}^s = [r_k, \dot r_k, a_k]^T, j = 1, \ldots, n_k^{e,s}$, which include slant range $r_k$, slant range rate $\dot r_k$ and azimuth $a_k$. In the presence of unknown ionospheric parameters, clutter and imperfect detection probability, the OTHR measurement function is
\begin{equation} \label{3-Measurement}
y_{j, k}^s =
\begin{cases}
    h_k\left(x_{i, k}, l^s, u_{1,k}^{s}\right) + w_{1,k}^{s}, &\text{if $y_{j,k}^s$ is the $i$th target-originated via path 1} \\
    \qquad \quad \vdots                              & \qquad \vdots \\
    h_k\Big(x_{i, k}, l^s, u_{n^m, k}^{s}\Big) + w_{n^m, k}^{s},  &\text{if $y_{j,k}^s$ is the $i$th target-originated via path $n^{m}$} \\
    \ c_k^{s}, \qquad \qquad \qquad  &\text{if $y_{j,k}^s$ is clutter}
\end{cases}
\end{equation}
where $h_k(\cdot)$ is the measurement function, $l^s = (x_0^s, y_0^s, \beta_0^s, d_0^s)$ is the configuration of OTHR $s$ consisting of receiver location $x_0^s$, $y_0^s$, bore-sight angle $\beta_0^s$, and distance $d_0^s$ between the receiver and the transmitter.
$u_{\tau,k}^{s}$ is the ionospheric height of the $\tau$th path, and $w_{\tau, k}^{s}$ is the zero-mean Gaussian white noise with known covariance $R_{\tau, k}^{s}$. Here, $v_{i,k}$, $w_{\tau, k}^{s}$ and $x_{i, 0}$ are assumed to be mutually independent.
The clutter $c_k^s$ is assumed uniformly distributed in the validation region with volume $V_k^s$, i.e., $p(c_k^s) = {1}/{V_k^s}$.

\begin{figure}[!htbp]
\centering
\includegraphics[scale = 0.7]{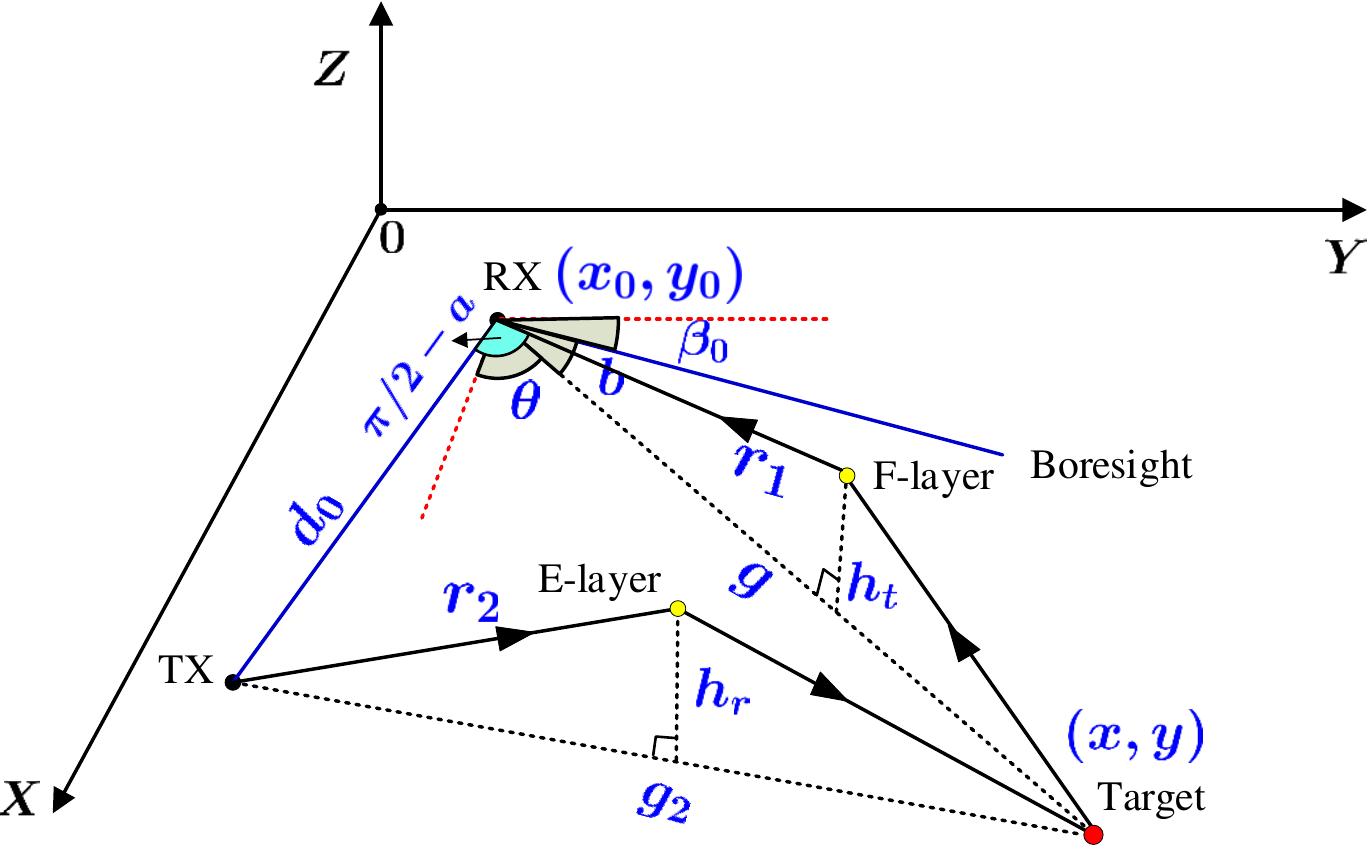}
\caption{Geometry of planar OTHR measurement model with target kinematic state being modeled in UTM coordinate system.
The position of the receiver~(RX) is at ($x_0, y_0$) and the distance between the transmitter~(TX) and the receiver is $d_0$.
The position of target is at $(x, y)$. EF propagation path is depicted. The corresponding measurement for the target includes slant range $r = r_1 + r_2$, azimuth $\pi / 2 - \theta$ and Doppler $\dot r$.}
\label{2-Scenario}
\end{figure}

From Fig.~\ref{2-Scenario}, the measurement function mapping the $i$th target state $[x_{i,k}, \dot x_{i,k}, y_{i,k}, \dot y_{i,k}]^T$ at time $k$ from UTM coordinate system to the $s$th radar slant coordinate system $[r_{k}^s, \dot r_k^s, a_k^s]^T$, i.e., $h_k(\cdot): \mathcal{MX} \times \mathcal{IH} \rightarrow \mathcal{RS}$, is expressible as~(for simplicity, the indices of notations, including time index $k$, target index $i$, and OTHR index $s$, are omitted.)
\begin{equation}\label{equ4}
\begin{split}
r =& \sqrt{\dfrac{g^2}{4} + h_r^2} + \sqrt{\dfrac{g^2 - 2d_0g\sin(b) + d_0^2}{4} + h_t^2}, \\
\dot r =& \dfrac{\dot g}{4} \left(\dfrac{g}{\sqrt{g^2 / 4 + h_r^2}} + \dfrac{g - d_0 \sin(b)}{\sqrt{\left(g^2 - 2d_0g\sin(b) + d_0^2\right) / 4 + h_t^2}}\right), \\
a =& \arcsin\left(\dfrac{g \sin(b)}{2\sqrt{g^2 / 4 + h_r^2}}\right),
\end{split}
\end{equation}
where
\begin{equation}
\begin{split}
g =& \sqrt{(x - x_0)^2 + (y - y_0)^2}, \\
\dot g =& \dfrac{(x-x_0)\dot x + (y-y_0)\dot y}{g}, \\
b =& \dfrac{\pi}{2} - \arctan\left(\dfrac{y - y_0}{x - x_0}\right) - \beta_0.
\end{split}
\end{equation}

The inverse mapping from radar slant coordinate system to UTM coordinate system, i.e., $\overline{h_k(\cdot)}: \mathcal{RS} \times \mathcal{IH} \rightarrow \mathcal{MX}$, required for track initiation, is given as
\begin{equation}\label{equ4}
\begin{split}
x =&~x_0 + \rho  \cos(\theta), \\
y =&~y_0 + \rho  \sin(\theta), \\
\end{split}
\end{equation}
where
\begin{equation}
\begin{split}
\rho =& 2 \sqrt{\left(\dfrac{r^2 + h_r^2 - h_t^2 - (d_0 / 2)^2}{2r - d_0 \sin(a)}\right)^2 - h_r^2}, \\
\theta =& \dfrac{\pi}{2} - \arcsin \left({2 \sin(a) \dfrac{r^2 + h_r^2 - h_t^2 - (d_0 / 2)^2}{\left(2r - d_0 \sin(a)\right)\rho}} \right) - \beta_0.
\end{split}
\end{equation}

The Jacobian matrices of the nonlinear measurement function $h_k(\cdot)$ with respect to~(w.~r.~t.) $x_{i,k}$ and $u_{k}^s$, are needed for target kinematic state estimation and ionospheric height identification, respectively.
The Jacobian matrix of $h_k(\cdot)$ w.~r.~t. $x_{i,k}$ is derived as follows.
\begin{equation}\label{eq:JXTau}
J_x^{\tau, s}(x_{i,k}, \hat u_{\tau,k}^{s}) =  \dfrac{\partial h_k\left(x_{i, k}, l^s, \hat u_{\tau,k}^{s}\right)}{\partial x_{i,k}} = \dfrac{\partial (r_k, \dot r_k, a_k)}{\partial (g_k, \dot g_k, b_k)} \times \dfrac{\partial (g_k, \dot g_k, b_k)}{\partial (x_k, \dot x_k, y_k, \dot y_k)},
\end{equation}
where
\begin{equation}
\begin{split}
&\dfrac{\partial (r_k, \dot r_k, a_k)}{\partial (g_k, \dot g_k, b_k)} \!=\! \begin{pmatrix}
                                                                         \dfrac{1}{4}\left(\dfrac{g}{r_1} + \dfrac{r_3}{r_2}\right) - \dfrac{d_0 \sin(b)}{4 r_2} & 0 & -\dfrac{d_0 g}{4 r_2} \cos(b) \\
                                                                          \dfrac{\dot g}{4}\left(\dfrac{1}{r_1} + \dfrac{1}{r_2} - \dfrac{g^2}{4r_1^3} - \dfrac{r_3^2}{4r_2^3} \right) & \dfrac{gr_2 + r_1r_3}{4r_1r_2} & \dfrac{-d_0\dot g}{2r_2}\left(1-\dfrac{gr_3}{8r_2^2}\right)\cos(b) \\
                                                                         \dfrac{\sin(b)(1 - g^2/4/r_1^2)}{2r_1\sqrt{1 - \left(\dfrac{g \sin(b)}{2r_1}\right)^2}} & 0 & \dfrac{g \cos(b)}{2 r_1 \sqrt{1 - \left(\dfrac{g \sin(b)}{2r_1}\right)^2}} \\
                                                                       \end{pmatrix}, \\
&\dfrac{\partial (g_k, \dot g_k, b_k)}{\partial (x_k, \dot x_k, y_k, \dot y_k)} = \begin{pmatrix}
                                                                                   \dfrac{(x - x_0)}{g} & 0 & \dfrac{(y - y_0)}{g} &  0 \\
                                                                                   \dfrac{\dot x}{g} - \dfrac{(x - x_0)r_4}{g^3} & \dfrac{x-x_0}{g} & \dfrac{\dot y}{g} - \dfrac{(y - y_0) r_4}{g^3}  & \dfrac{y-y_0}{g} \\
                                                                                   -\dfrac{(y - y_0) }{g^2} & 0 & \dfrac{(x - x_0)}{g^2} & 0 \\
                                                                                 \end{pmatrix}
\end{split}
\end{equation}
with
\begin{equation}
\begin{split}
r_1 =& \sqrt{g^2 / 4 + h_r^2}, \\
r_2 =& \sqrt{\left(g^2 - 2d_0g\sin(b) + d_0^2\right) / 4 + h_t^2}, \\
r_3 =& g - d_0 \sin(b), \\
r_4 =& (x-x_0)\dot x + (y-y_0)\dot y.
\end{split}
\end{equation}

The Jacobian matrix of $h_k(\cdot)$ w.~r.~t. $u_{k}^{m, s}$ is derived as follows.
\begin{equation}\label{eq:JUTauS}
\begin{split}
J_u^{\tau, s}(\hat x_{i,k}, u_{\tau,k}^{s}) =& \dfrac{\partial h_k\left(\hat x_{i, k}, l^s,  u_{\tau,k}^{s}\right)}{\partial u_{k}^{\tau, s}}
                                     = \begin{pmatrix}
                                          \dfrac{h_t}{r_2} & \dfrac{h_r}{r_1} \\
                                          \dfrac{d_0 h_t \dot g \sin(b)}{4 r_2^3} & \dfrac{-g \dot g h_t }{4 r_1^3} \\
                                          0 & \dfrac{-g h_r \sin(b)}{2r_1^3 \sqrt{1 - \dfrac{(g \sin(b))^2}{g^2 + 4 h_r^2}}} \\
                                        \end{pmatrix}.
\end{split}
\end{equation}

\subsection{Multipath Data Association Modeling}
To reduce the computational cost of multipath data association,
we assume that the measurements from different OTHRs are associated with a target individually, i.e., $A_k = \prod_{s = 1}^{n^s} A_k^s$.
Let $A_k^s = \big\{a^{s}_{i, j, \tau, k}\big\}_{i = 1, \cdots, n_k^x, j = 0, \cdots, n_k^{e,s}, \tau = 1,\cdots, n^m} \bigcup \big\{a^{s}_{0,j,k}\big\}_{j=1, \cdots, n_k^{e,s}}$ be the joint multipath data association event of OTHR $s$ at time $k$.
The binary association variable $a^{s}_{i,j,\tau,k} \in \{0, 1\}$ represents an association event of target-to-measurement-to-path.
In particular, $a^{s}_{i,j,\tau,k}(i > 0, j >0)$ represents that the $j$th measurement is originated from the $i$th target via the $\tau$th path,
$a^{s}_{i,0,\tau,k}(i > 0, j = 0)$ represents that the $i$th target is missed by path $\tau$, and $a^{s}_{0, j, k}(i = 0, j > 0)$ represents that measurement $j$ is clutter where the index $\tau$ is dropped by the fact that clutter is irrespective of a propagation path.
In OTHR target tracking, all the feasible joint multipath data association event are constructed according to the following two assumptions:
(1) at each time, a measurement is either originated from one target via a particular path or it is clutter; (2) at each time, under a particular path, each target generates at most one measurement. Based on the above assumptions, the association variable $a^{s}_{i,j,\tau,k}$ should fulfill the following equations,
\begin{equation}\label{AssHpo}
\begin{split}
&\sum_{i=1}^{n_k^x} \sum_{\tau=1}^{n^{m}} a^{s}_{i,j,\tau,k} + a^{s}_{0,j,k} = 1, \quad   j = 1, \ldots, n_k^{e,s}, \, \forall k, \, \forall s,\\
&\sum_{j=0}^{n_k^{e,s}} a^s_{i,j,\tau,k} = 1, \quad \forall i = 1, \ldots, n_k^x, \;\;  \tau = 1, \ldots, n^{m}, \, \forall k, \, \forall s.
\end{split}
\end{equation}
We call the equations in~(\ref{AssHpo}) as one-to-one (hard) frame constraints. A joint association event $A_k^s$ is called \emph{feasible} if it fulfills the frame constraints, i.e., $\mathbb{I}(A_k^s \in \mathcal{A}_k^s)$ with $\mathcal{A}_k^s$ being the set of all feasible joint association events.

Given $n_k^x$ targets, $n_k^{e,s}$ measurements, $n^{m}$ paths, and by assuming that the number of clutter is Possion distributed with density $\lambda^s$, the prior PDF of multipath data association $A^s_k$ conditioned on the target visibility state $E_k$ is~\cite{Lan2015IRS}
\begin{equation}\label{proA}
p(A^s_k|E_k) = \dfrac{(\lambda^s V_k^s)^{n_k^c}}{n_k^{e,s}!}\exp(-\lambda^s V_k^s)\prod_{i=1}^{n_k^x}\prod_{\tau=1}^{n^{m}}\left(P_d^{\tau,s}(e_{i,k})\right)^{d_k^{i, \tau}} \left(1-P_d^{\tau,s}(e_{i,k})\right)^{1-d_k^{i, \tau}},
\end{equation}
where $n_k^c = n_k^{e,s} - \sum_{i=1}^{n_k^x}\sum_{\tau=1}^{n^{m}}d_k^{i,\tau}$ is the number of unassociated measurements~(clutter) at time $k$ in $A_k^s$, and $d_k^{i,\tau} = 1 - a^{s}_{i,0,\tau,k}$ is the path-dependent target detection indicator.
The two-valued variable $P_d^{\tau,s}(e_{i,k})$, i.e., $P_d^{\tau,s}(e_{i,k} = 1) = p_d^{\tau,s}$ and $P_d^{\tau,s}(e_{i,k} = 0) = \varepsilon~(0 < \varepsilon \ll 1)$, represents the target visibility state-dependent detection probability.

\subsection{Problem Statement}
Let the joint latent variables $\Theta_{1:K} = \{E_{1:K}, X_{1:K}, A_{1:K}, U_{1:K}\}$ with $E_{1:K}$, $X_{1:K}$, $A_{1:K}$, $U_{1:K}$ being the sequences from time 1 to $K$ of target visibility state, target kinematic state, multipath data association, ionospheric height, respectively. Denote the joint observation variables $Y_{1:K}$ and $I_{1:K}$ as the sequences of radar measurements and ionospheric measurements from time 1 to $K$ for all OTHRs, respectively.
The task of multitarget tracking of the OTHR network is to perform target detection~$E_{1:K}$, multipath data association~$A_{1:K}$, ionospheric heights identification~$U_{1:K}$ and target tracking~$X_{1:K}$ simultaneously, given measurements~$Y_{1:K}$ and $I_{1:K}$.
In the optimal Bayesian framework, it is required to solve the joint posterior PDFs $\mathcal{L}(\Theta_{1:K}) = p\left(\Theta_{1:K}|Y_{1:K}, I_{1:K}\right)$ first, and then marginalize it to obtain the posterior PDFs of each latent variables.
The interdependent relationships among the latent variables are assumed as follows.
(1) The global latent variables~(independent with local OTHR), i.e., target visibility state $E_{1:K}$ and target kinematic state $X_{1:K}$, evolve with first-order Markov process.
(2) The local latent variables~(dependent with local OTHR) include multipath data association $A_{1:K}^s$ and ionospheric height $U_{1:K}^s$, $s = 1, \ldots, n^s$, whereas $A_{1:K}^s$ is independent over time and $U_{1:K}^s$ evolves with first-order Markov process.
(3) Given $X_k$, $A_k^s$ and $u_k^s$, the measurement $Y_k$ are conditionally independent across $s$ and $\tau$.
(4) Given $u_k^s$, the ionosphere measurements $I_k$ are conditionally independent across $s$.
(5) $A_k$ is related to target visibility state $E_k$.
(6) Given $A_k$, $X_k$ is conditionally independent of $E_k$ .
(7) Global latent variables $X_{1:K}, E_{1:K}$ can be factorized over targets~(i.e., targets are assumed to be independent).
Based on the above assumptions, the factorization of the full joint posterior PDFs $\mathcal{L}(\Theta_{1:K})$ is given as
\begin{equation} \label{5-joint-posterior}
\begin{split}
\mathcal{L}\left(\Theta_{1:K}\right) \propto \prod_{s = 1}^{n^s} \prod_{k = 1}^K \prod_{j=1}^{n_k^{e, s}} \dfrac{1}{V_k^s} \prod_{i = 1}^{n_k^x} p\left(y_{j,k}^{s}|x_{i,k}, u_k^s, A_k^s\right)p\left(x_{i,k}|x_{i,k-1}\right) \times \prod_{s = 1}^{n^s} \prod_{k = 1}^K p\left(I_k^s|u_k^s\right) p\left(u_k^s|u_{k-1}^s\right) \\
\times \prod_{s = 1}^{n^s} \prod_{k = 1}^K \mathbb{I}\left(A_k^s \in \mathcal{A}_k^s\right) p\left(A_k^s|E_k\right) \times \prod_{k = 1}^K \prod_{i = 1}^{n_k^x} p\left(e_{i,k}|e_{i,k-1}\right).
\end{split}
\end{equation}

The following requirements must be considered when solving the joint inference problem (\ref{5-joint-posterior}).
\begin{itemize}
\item Marginalize the high-dimensional joint posterior PDFs $\mathcal{L}(\Theta_{1: K})$ in Eq.~(\ref{5-joint-posterior}) is intractable because the required integration w.~r.~t. continuous latent variables~($X_{1:K}$, $U_{1:K}$) may not have closed-form analytical solutions due to the nonlinear transformation among the state $X_{1:K}$, measurements $Y_{1:K}$ and height $U_{1:K}$, and the required summation w.~r.~t. discrete latent variables~($A_{1:K}$, $E_{1:K}$) is prohibitively expensive due to the complex multipath data association. In such case, an approximate Bayesian approach is preferable.
\item There exists correlation among target state estimation and environmental parameters identification. That is, the target state estimation~($X_{1:K}$, $E_{1:K}$) depends on the identification of parameters ($A_{1:K}^s$, $U_{1:K}^s$), and can be used to improve the identification of the parameters.
    A joint solution is especially important because the solution of one can greatly help improve the other.
    As above-stated, the optimal solution is usually intractable, which can be approximated by an iterative optimization approach.
\item There exist both global latent variables $X_{1:K}$, $E_{1:K}$ and local latent variables $A_{1:K}^s$, $U_{1:K}^s$, $s=1, \ldots, n^s$.
It is demanded to design a two-layer~(global-local) processing structure, whereas information is exchanged in both directions between the two layers. The multipath measurements from multiple OTHRs $Y_1^k$ are integrated to estimate the global latent variables by using the identified local latent variables $A_{1:K}^s$ and $U_{1:K}^s$, and the updated global latent variables are feedback to improve the estimation of the local latent variables.
\end{itemize}
	
\section{MP-Based Multisensor Multipath Measurement-level Fusion Approach}\label{sec:solutions}
\subsection{General Framework}
\begin{figure}[!htbp]
\centering
\includegraphics[scale = 0.5]{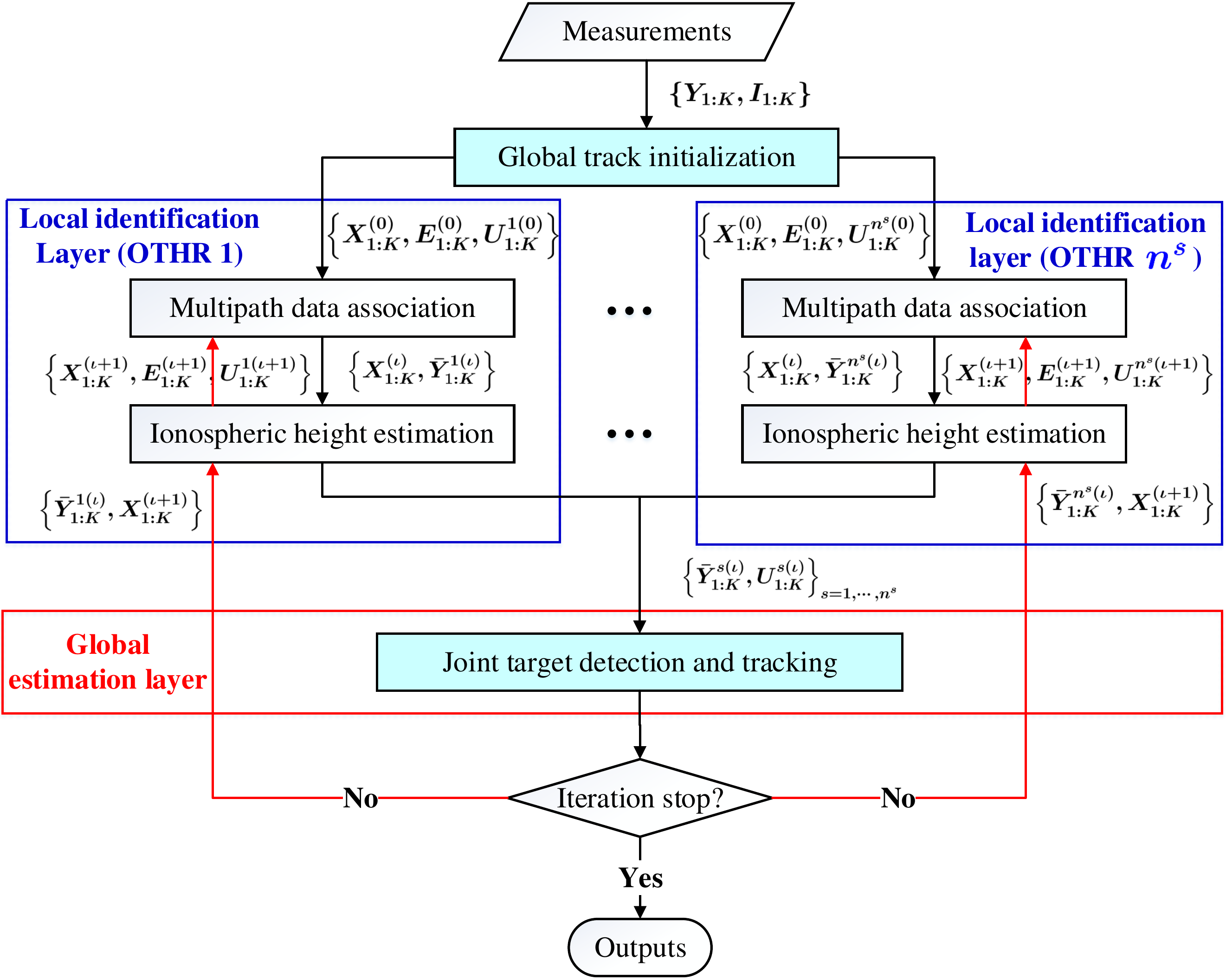}
\caption{The diagram of MP-OTHRs with a two-layer processing structure.}
\label{1-Flowchart}
\end{figure}

The proposed MP-OTHRs, is an iterative and joint target detection, multipath data association, ionospheric heights estimation and target tracking solution with a two-layer processing structure including a local identification layer and a global estimation layer.
The diagram of MP-OTHRs is depicted in Fig.~\ref{1-Flowchart}, which is further explained as follows.

Considering a time sequence $k = 1, \cdots, K$, set the initial target kinematic state $X_{1:K}^{(0)}$ and target visibility state $E_{1:K}^{(0)}$ via global track initialization process.
At the $\iota$th iteration, for each OTHR $s$, $s = 1, \cdots, n^s$,
the local identification layer associates the measurements with the underlying targets and the propagation path, i.e., multipath data association $A_{1:K}^{s(\iota)}$, resulting in the path-dependent pseudo-measurements of each target $\bar Y^{s(\iota)}_{1:K} = \big\{\bar y_k^{i, \tau, s}\big\}_{i = 1, \cdots, n_k^x, \tau = 1, \cdots, n^m, k = 1, \cdots, K}$,
and estimates the ionospheric height $U_{1:K}^{s(\iota)}$ by using the global kinematic state and its corresponding pseudo-measurements.
Such path-dependent pseudo-measurements together with the corresponding estimated ionospheric height are then integrated to update the global kinematic state $X^{(\iota)}_{1:K}$ and visibility state $E^{(\iota)}_{1:K}$.
The information is exchanged between the local identification layer and the global estimation layer until convergence.
Compared with the track-level fusion framework without feedback, the performance of the local identification layer can be improved since it uses the global estimation results rather than local ones.
The global estimation layer carries out the target track detection and tracking based on all pseudo-measurements and estimated ionospheric height, benefiting to improving the tracking performance.
The next sections detail the proposed MP-OTHRs approach in a unified MP framework, which combines MF and BP based on the region-based free energy approximation.

\subsection{Combined BP-MF Approximation for OTHR Network Fusion}
\begin{figure}[!htbp]
\centering
\includegraphics[scale = 0.55]{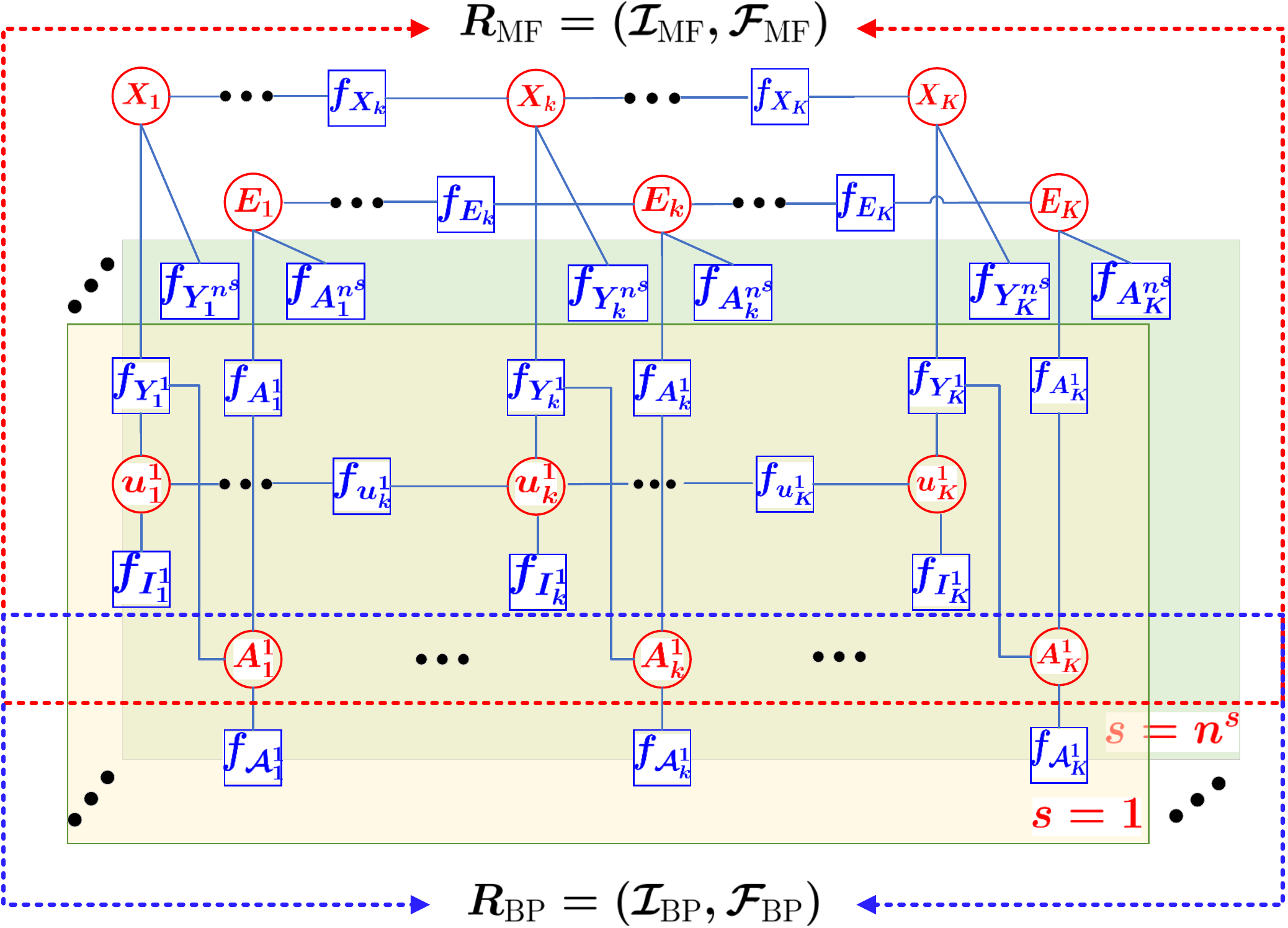}
\caption{Factor graph of Eq.~(\ref{5-joint-posterior}) for OTHR network target tracking problem, which is partitioned into BP region and MF region by the fact that MF is suitable for conjugate-exponential models and BP works well with hard constraints. For simplicity, the following short notations of factor nodes are used: $ f_{X_k} \triangleq \prod_{i=1}^{n_k^s} p(x_{i,k}|x_{i,k-1}), f_{E_k} \triangleq \prod_{i=1}^{n_k^s} p(e_{i,k}|e_{i,k-1}), f_{u^s_k} \triangleq p(u_k^s|u_{k-1}^s), f_{Y^s_k} \triangleq p(Y_k^s|X_k, A_k^s, u_k^s), f_{I^s_k} \triangleq p(I_k^s|u_k^s), f_{A^s_{k}} \triangleq p(A_k^s|E_k),f_{\mathcal{A}_k^s} \triangleq \mathbb{I}(A_k^s \in \mathcal{A}_k^s)$.}
\label{2-FG}
\end{figure}

As aforementioned, the difficulty of the measurement-level fusion for the OTHR network arises from solving the intractable joint posterior PDF $\mathcal{L}(\Theta_{1:K})$ in Eq.~(\ref{5-joint-posterior}).
This intractable inference can be approximated by running the message passing on a factor graph. In particular, the factor graph model of the factorization of Eq.~(\ref{5-joint-posterior}) is illustrated as Fig.~\ref{2-FG}, which consists of \emph{variable node} $i \in \mathcal{I}$ ~(represented by a red circle) for each variable $x_i$, \emph{factor node} $\alpha \in \mathcal{F}$~(represented by a blue square) for each local function $f_{\alpha}$, and an edge connecting variable node $i$ to factor node $\alpha$ if and only if $x_i$ is an argument of $f_{\alpha}$, where $\mathcal{I}$ and $\mathcal{F}$ are the sets of all variable nodes and factor nodes, respectively.

Following the definitions in~\cite{yedidia2005constructing, riegler2012merging}, a \emph{region} $R \triangleq \{\mathcal{I}_R, \mathcal{F}_R\}$ are subsets of variable nodes $\mathcal{I}_R \subset \mathcal{I}$ and factor nodes $\mathcal{F}_R \subset \mathcal{F}$ in a factor graph such that if a factor node $\alpha$ belongs to $\mathcal{F}_R$, all the variable nodes neighboring $a$ are in $\mathcal{I}_R$.
Each region $R$ associates a \emph{counting number} $c_R \in \mathbb{Z}$. We say a set $\mathcal{R} = \{(R, c_R)\}$ of regions and associated counting numbers gives a \emph{valid} region-based approximation if
\begin{equation}
\sum_{(R, c_R) \in \mathcal{R}} c_R \mathbb{I}(\alpha \in \mathcal{F}_R) = \sum_{(R, c_R) \in \mathcal{R}} c_R\mathbb{I}(i \in \mathcal{I}_R) = 1.
\end{equation}

As shown in Fig.~\ref{2-FG}, the factor factor can be divided into two regions, i.e., MF region $R_{\text{MF}} = (\mathcal{I}_{\text{MF}}, \mathcal{F}_{\text{MF}})$ and BP region $R_{\text{BP}} = (\mathcal{I}_{\text{BP}}, \mathcal{F}_{\text{BP}})$ with
\begin{equation}
\begin{split}
\mathcal{I}_{\text{MF}} =& \big\{X_k \cup E_k\big\}_{k = 1, \ldots,K} \cup \big\{U_k^s \cup A_k^s \big\}_{k = 1, \ldots, s=1, \ldots, n^s},\\
\mathcal{F}_{\text{MF}} =& \big\{f_{X_{k}} \cup f_{E_{k}}\big\}_{k = 1,\ldots, K}  \cup  \big\{f_{u^s_{k}} \cup f_{Y^s_{k}} \cup f_{I^s_{k}} \cup f_{A^s_{k}}\big\}_{k = 1, \ldots, K, s=1, \ldots, n^s}, \\
\mathcal{I}_{\text{BP}} =& \big\{A_k^s\big\}_{k = 1, \ldots, K, s=1,\ldots, n^s}, \\
\mathcal{F}_{\text{BP}} =& \big\{f_{\mathcal{A}_k^s}\big\}_{k = 1, \ldots, K, s=1, \ldots, n^s}
\end{split}
\end{equation}

It is seen that $\mathcal{I}_{\text{BP}} \cup \mathcal{I}_{\text{MF}} = \mathcal{I}$, $\mathcal{I}_{\text{BP}} \cap \mathcal{I}_{\text{MF}} = \Big\{A^1_1, \ldots, A^1_K\Big\} \cup \cdots \cup \Big\{A^{n^s}_1, \ldots, A^{n^s}_K\Big\}$, $\mathcal{F}_{\text{BP}} \cup \mathcal{F}_{\text{MF}} = \mathcal{F}$ and $\mathcal{F}_{\text{BP}} \cap \mathcal{F}_{\text{MF}} = \emptyset$.
According to~\cite{riegler2012merging}, the joint posterior PDF $\mathcal{L}(\Theta_{1:K})$ is expressed as
\begin{equation}\label{RFG}
\mathcal{L}(\Theta_{1:K}) = \overbrace{f_{X_{1:K}} \times f_{E_{1:K}} \times f_{u^1_{1:K}}\times \!\cdots\! \times f_{u^{n^s}_{1:K}} \times f_{A^1_{1:K}} \times \!\cdots\! \times f_{A^{n^s}_{1:K}}}^{\text{MF region}} \times \overbrace{f_{\mathcal{A}_{1:K}^1} \times \!\cdots\! \times f_{\mathcal{A}_{1:K}^{n^s}}}^{\text{BP region}}.
\end{equation}
An approximation of marginal PDFs of each variables in $\mathcal{L}(\Theta_{1:K})$ can be derived by  
minimizing the \emph{region-based free energy} $F_{\text{BP, MF}}$, which is defined by~\cite{yedidia2005constructing}
\begin{equation}
\begin{split}
F_{\text{BP, MF}} =& \sum_{\alpha \in \mathcal{F}_{\text{BP}}}\sum_{\bm{x}_{\alpha}} b_{\alpha}(\bm{x}_{\alpha}) \ln \dfrac{b_{\alpha}(\bm{x}_{\alpha})}{f_{\alpha}(\bm{x}_{\alpha})} - \sum_{\alpha \in \mathcal{F}_{\text{MF}}}\sum_{\bm{x}_{\alpha}} \prod_{i \in \mathcal{S}(\alpha)}b_i(x_i) \ln f_{\alpha}(\bm{x}_{\alpha}) \\
&- \sum_{i \in \mathcal{I}}(|\mathcal{S}_{\text{BP}}(i) - 1|) \sum_{x_i}b_i(x_i)\ln b_i(x_i),
\end{split}
\end{equation}
where $\bm{x}_{\alpha} \triangleq \left(x_i | i \in \mathcal{S}(\alpha)\right)^T$, and the positive functions $b_{\alpha}(\bm{x}_{\alpha})$ and $b_i(x_i)$, referred as beliefs, are the approximations of $f_{\alpha}(\bm{x}_{\alpha})$ and $p(x_i)$, respectively.
The beliefs $b_{\alpha}(\bm{x}_{\alpha})$ and $b_i(x_i)$ have to fulfill the marginalization constraints
\begin{equation}\label{eq:marconstraint}
b_i(x_i) = \sum_{\bm{x}_{\alpha} \backslash x_i} b_{\alpha}(\bm{x}_{\alpha}), \quad \forall \alpha \in \mathcal{F}_{\text{BP}}, i \in \mathcal{S}(\alpha),
\end{equation}
and the normalization constraints
\begin{equation}\label{eq:norconstraint}
\begin{split}
&\sum_{x_i} b_i(x_i) = 1, \quad \forall i \in \mathcal{I}_{\text{MF}} \backslash \mathcal{I}_{\text{BP}},\quad \sum_{\bm{x}_{\alpha}} b_{\alpha}(\bm{x}_{\alpha}) = 1,  \quad \forall {\alpha} \in \mathcal{F}_{\text{BP}}.
\end{split}
\end{equation}
Using the Lagrange multipliers method with the constraints given in Eqs.~(\ref{eq:marconstraint}) and (\ref{eq:norconstraint}),
the combined BP-MF approach~\cite{riegler2012merging} yields the belief $b_i(x_i)$, i.e., the approximation to the exact marginal probability function $p_i(x_i)$, as follows.
\begin{equation}\label{eq:bx}
b_i(x_i) = z_i \prod_{{\alpha} \in \mathcal{S}_{\text{BP}}(i)} m_{{\alpha} \rightarrow i}^{\text{BP}}(x_i) \prod_{{\alpha} \in \mathcal{S}_{\text{MF}}(i)} m_{{\alpha} \rightarrow i}^{\text{MF}}(x_i), \quad \forall i \in \mathcal{I}.
\end{equation}
with the message update rules given by
\begin{equation}\label{Eq-update}
\begin{split}
n_{i \rightarrow {\alpha}}(x_i) =& z_i \prod_{c \in \mathcal{S}_{\text{BP}}(i) \backslash {\alpha}} m_{c \rightarrow i}^{\text{BP}}(x_i) \prod_{c \in \mathcal{S}_{\text{MF}}(i)} m_{c \rightarrow i}^{\text{MF}}(x_i), \quad \forall a \in \mathcal{F}, i \in \mathcal{S}({\alpha}) \\
    m_{{\alpha} \rightarrow i}^{\text{BP}}(x_i) =& z_{\alpha} \sum_{\bm{x}_{\alpha} \backslash x_i} f_{\alpha}(\bm{x}_{\alpha}) \prod_{j \in \mathcal{S}({\alpha}) \backslash i} n_{j \rightarrow {\alpha}}(x_j), \quad \quad \quad  \forall {\alpha} \in \mathcal{F_{\text{BP}}}, i \in \mathcal{S}({\alpha}) \\
    m_{{\alpha} \rightarrow i}^{\text{MF}}(x_i) =& \exp\Bigg( \sum_{\bm{x}_{\alpha} \backslash x_i} \prod_{j \in \mathcal{S}({\alpha}) \backslash i} n_{j \rightarrow {\alpha}}(x_j) \ln  f_{\alpha}(\bm{x}_{\alpha})  \Bigg), \;\; \forall {\alpha} \in \mathcal{F_{\text{MF}}}, i \in \mathcal{S}({\alpha})
\end{split}
\end{equation}
where $n_{i \rightarrow {\alpha}}(x_i)$ denotes the message sent from variable node $i$ to factor node ${\alpha}$,
and $m_{{\alpha} \rightarrow i}(x_i)$ denotes the message sent from factor node ${\alpha}$ to variable node $i$.
The notation $\mathcal{S}({\alpha}) \backslash i$ denotes the set of variable nodes $\mathcal{S}({\alpha})$ except variable node $i$, and $\sum_{\bm{x}_{\alpha} \backslash x_i}$ denotes a sum over all the variables $\bm{x}_{\alpha}$ except $x_i$. $z_i~(i \in \mathcal{I})$ and $z_{\alpha} ({\alpha} \in \mathcal{F}_{\text{BP}})$ are positive constants ensuring normalized beliefs.
Note that $n_{i \rightarrow {\alpha}}(x_i) = b_i(x_i)$ when ${\alpha} \in \mathcal{F}_{\text{MF}}$.

The detailed derivations of each belief with the corresponding subgraphs of the factor graph are presented in the remainder of this section.
Note that MP-OTHRs works in a batch processing fashion.
For a batch time sequence $[1, K]$, different targets may have a different lifetime since target may appear/disapper anytime and anywhere in the region of interests. For ease of exposition~(and without loss of generality), we hereafter restrict the notation on the lifetime of all targets from 1 to $K$, and the maximum number of targets during $[1, K]$ is $n^x$.

\subsubsection{Derivation of Belief~$b_X(X)$}
By the fact that each target moves independently, the belief of the joint kinematic states of all targets can be factorized as
\begin{equation}\label{eq:bKS}
b_X(X) = \prod_{i = 1}^{n^x}b_X\left(x_{i,1:K}\right) = \prod_{i = 1}^{n^x} \prod_{k=1}^K b_X(x_{i,k}).
\end{equation}
\begin{figure}[!htp]
\centering
\includegraphics[width=0.55\textwidth]{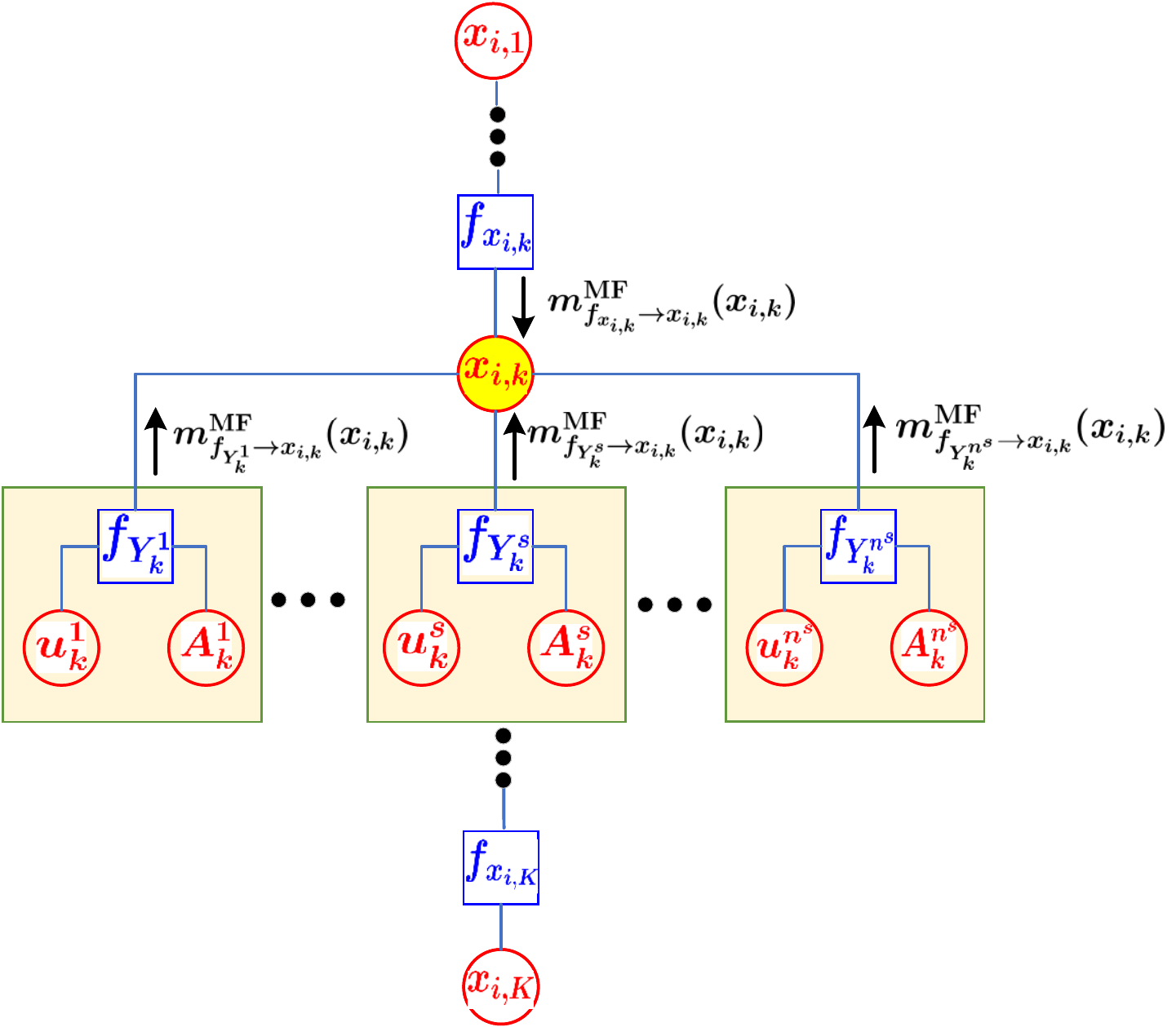}
\caption{The target kinematic state estimation subgraph of $b_X(x_{i,k})$.}
\label{Subgraph-x}
\end{figure}
Fig.~\ref{Subgraph-x} shows the target kinematic state estimation subgraph of the belief $b_X(x_{i,k})$.
In Fig.~\ref{Subgraph-x}, $x_{i,k}, i = 1, \ldots, n^x, k = 1, \ldots, K$, are the variable nodes to be considered, and our aim is to calculate belief $b_X\left(x_{i,k}\right)$.
For each variable node $x_{i,k}$, $\mathcal{S}\left(x_{i,k}\right) = \big\{f_{x_{i,k}}, f_{Y_k^1}, \ldots, f_{Y_k^{n^s}}\big\}$ is the set of all factor nodes connecting to the variable node $x_{i,k}$. Meanwhile, the sets of variable nodes connected to the each factor node in $\mathcal{S}\left(x_{i,k}\right)$ are $\mathcal{S}\left(f_{x_{i,k}}\right) = \left\{x_{i,k}, x_{i,k-1}\right\}$ and $\mathcal{S}\left(f_{Y_k^s}\right) = \left\{x_{i,k}, A_k^s, u_k^s\right\}, s = 1, \ldots, n^s$, respectively.
According to Eq.~(\ref{eq:bx}), the belief $b_X(x_{i,k})$ can be calculated as follows
\begin{align}\label{eq:bKSx}
b_X\left(x_{i,k}\right) \propto
    m^{\text{MF}}_{f_{x_{i,k}} \rightarrow x_{i,k}}\left(x_{i,k}\right) \times \prod_{s=1}^{n^s} m^{\text{MF}}_{f_{Y_k^s} \rightarrow x_{i,k}}\left(x_{i,k}\right).
\end{align}
where the factor-to-variable messages in Eq.~(\ref{eq:bKSx}) can be calculated as follows by using the message update rules given in Eq.~(\ref{Eq-update})
\begin{align}
&m^{\text{MF}}_{f_{x_{i,k}} \rightarrow x_{i,k}}(x_{i,k}) = \exp\left(\int_{x_{i,k-1}}  n_{x_{i,k-1} \rightarrow f_{x_{i,k}}}\left(x_{i,k-1}\right)\ln p\left(x_{i,k}|x_{i,k-1}\right) d_{x_{i, k-1}} \right) \label{eq:F2VM}, \\
&m^{\text{MF}}_{f_{Y_k^s} \rightarrow x_k^i}(x_{i,k}) \!=\! \exp \left(\!\int_{u_k^s}\! \sum_{j = 1}^{n_k^{e,s}} \!\sum_{\tau = 1}^{n^{m}} n_{a_{i,j,\tau,k}^s \rightarrow f_{Y_k^s}}\!\!\left(a_{i,j,\tau,k}^s\right)\! n_{u_k^s \rightarrow f_{Y_k^s}}\!\left(u_k^s\right)\! \ln p\left(y^s_{j,k}|x_{i,k}, \!u_k^s, \!a_{i,j,\tau,k}^s\right) \!d_{u_k^s} \right) \label{eq:F2VM2}.
\end{align}
Recall that the variable-to-factor messages $n_{i \rightarrow a}(x_i) = b_i(x_i)$, $\forall a \in \mathcal{F}_{\text{MF}}$ and $i \in \mathcal{S}(a)$. Thus,
\begin{equation}
\begin{split}
n_{x_{i,k-1} \rightarrow f_{x_{i,k}}}\left(x_{i,k-1}\right) =&~ b_X\left(x_{i,k-1}\right), \\
n_{a_{i,j,\tau,k}^s \rightarrow f_{Y_k^s}}\left(a_{i,j,\tau,k}^s\right) =&~ b_A\left(a_{i,j,\tau,k}^s\right), \\
n_{u_k^s \rightarrow f_{Y_k^s}}\left(u_k^s\right) =&~ b_U\left(u^s_k\right) \label{eq:V2FM2}.
\end{split}
\end{equation}

Substituting Eq.~(\ref{eq:V2FM2}) into Eqs.~(\ref{eq:F2VM}), (\ref{eq:F2VM2}), yields
\begin{align}
m^{\text{MF}}_{f_{x_{i,k}} \rightarrow x_{i,k}}(x_{i,k}) =& \mathcal{N}\left(x_{i,k}|F_k\hat x_{i,k-1}, F_k P_{i,k-1}F_k^T + Q_k\right) \label{eq:mMFpx2x}, \\
m^{\text{MF}}_{f_{Y_k^s} \rightarrow x_k^i}(x_{i,k}) =& \prod_{j = 1}^{n_k^{e,s}} \prod_{\tau = 1}^{n^{m}}p\left(y^s_{j,k}|x_{i,k}, \hat u_{\tau,k}^{s} \right)^{\hat a_{i,j,\tau,k}^s} \label{eq:mMFpx2x2},
\end{align}
where $\hat x_{i, k-1} =\langle x_{i, k-1} \rangle_{b_X(x_{i,k-1})}$, $\hat a_{i,j,\tau,k}^s = \langle a_{i,j,\tau,k}^s \rangle_{b_A(a_{i,j,\tau,k}^s)}$ and  $\hat u_{\tau,k}^{s} =\langle u_{\tau,k}^{s} \rangle_{b_U(u_{\tau,k}^{s})}$ are the expectations of $x_{i,k-1}$, $a_{i,j,\tau,k}^s$ and $u_{\tau,k}^{s}$ taken over corresponding beliefs, respectively.
$P_{i,k-1}$ is the state estimation covariance of target $i$ at time $k-1$.

For Gaussian-distributed ionospheric height $u_{\tau,k}^{s}$ under a given propagation path $\tau$, the measurement distribution $p\left(y^s_{j,k}|x_{i,k}, \hat u_{\tau,k}^{s} \right)$ in Eq.~(\ref{eq:mMFpx2x2}) under the nonlinear mapping $h_k(x_{i,k}, l^s, \hat u_{\tau,k}^{s})$ of a given target state $x_{i,k}$ is in general non-Gaussian.
In the vein of~\cite{pulford2004othr}, we use Gaussian approximations, i.e.,
\begin{equation}\label{Mlinear}
p\left(y^s_{j,k}|x_{i,k}, \hat u_{\tau,k}^{s} \right) \approx \mathcal{N}\left(y_{j,k}^s|h_k\left(x_{i,k}, l^s, \hat u_{\tau,k}^{s}\right), R_{i, \tau, k}^s\right),
\end{equation}
where the covariance $R_{i,\tau,k}^s$ is expressed as the sum of two components $R_{i,\tau,k}^s = R_{\tau,k}^{s} + R_{u, \tau,k}^{s}$,
including the measurement noise covariance $R_{\tau,k}^{s}$ of OTHR $s$ and the covariance $R_{u, \tau,k}^{s}$ arising from the uncertain ionospheric height.
The latter component is evaluated to a first-order approximation as $R_{u, \tau,k}^{s} = J_u^{\tau, s} \Sigma_{\tau, k}^{s} (J_u^{\tau, s})^T$ with $J_u^{\tau, s}$ being given by Eq.~(\ref{eq:JUTauS}), and $\Sigma_{\tau, k}^{s}$ being the state covariance of the ionospheric height at time $k$.

Substituting Eqs.~(\ref{eq:mMFpx2x}), (\ref{eq:mMFpx2x2}) and (\ref{Mlinear}) into Eq.~(\ref{eq:bKSx}), the belief $b_X(x_{i,k})$ is rewritten as
\begin{equation}\label{eq:bKSxUpdate}
\begin{split}
b_X(x_{i,k}) \propto \mathcal{N}\left(x_{i,k}|F_k\hat x_{i,k-1}, F_k P_{i,k-1}F_k^T + Q_k\right)\prod_{s=1}^{n^s}\prod_{\tau = 1}^{n_k^{m,s}}
\mathcal{N}\left(\bar y_{i, \tau, k}^{s}|h_k\big(x_{i,k}, l^s, \hat u_{\tau,k}^{s} \big), \bar R_{i, \tau, k}^{s}\right),
\end{split}
\end{equation}
where the synthetic measurement $\bar y_{i,\tau, k}^{s}$ and the corresponding covariance $\bar R_{i,\tau,k}^{s}$ are defined as
\begin{equation} \label{App-bar-y}
\bar{y}_{i,\tau, k}^{s} =  \dfrac{\sum \nolimits_{j=1}^{n_k^{e,s}} \hat a^{s}_{i,j, \tau, k}y_{j, k}^s}{1- \hat a^{s}_{i,0, \tau, k}},\quad
\bar{R}_{i, \tau, k}^{s} = \frac{R_{i, \tau, k}^{s}}{1- \hat a^{s}_{i,0, \tau, k}}.
\end{equation}

Since for different OTHR $s$ and propagation path $\tau$, the measurement function $h_k(\cdot)$ and the ionospheric height $u^s_{\tau,k}$ are different, the synthetic measurements $\bar y_{i, \tau, k}^{s}$ cannot be synthesized further over $\tau$ and $s$.
Let
\begin{equation}
    \textbf{y}_{i,k} = \begin{bmatrix}
    \bar y_{i,1,k}^{1} \\
    \vdots \\
    \bar y_{i,n^m,k}^{n^s}
    \end{bmatrix}, \quad
    \textbf{h}_{i,k} = \begin{bmatrix}
    h_k\Big(x_{i,k}, l^{1}, \hat u_{1,k}^{1}\Big) \\
    \vdots \\
    h_k\Big(x_{i,k}, l^s, \hat u_{n^m, k}^{n^s}\Big)
    \end{bmatrix}, \quad
    \textbf{R}_{i,k} =  \begin{bmatrix}
    \bar R_{i,1, k}^{1} & & \\
    & \ddots & \\
    &  & \bar R_{i,n^m,k}^{n^s}
    \end{bmatrix},
\end{equation}
The belief $b_X(x_{i,k})$ in Eq.~(\ref{eq:bKSxUpdate}) can be rewritten as
\begin{equation}\label{eq:bKSxUpdate1}
    b_X(x_{i,k}) \propto \mathcal{N}\left(x_{i,k}|F_k\hat x_{i,k-1}, F_k P_{i,k-1}F_k^T + Q_k\right) \mathcal{N}\left(\textbf{y}_{i,k}|\textbf{h}_{i,k}, \textbf{R}_{i,k}\right).
\end{equation}

From Eq.~(\ref{eq:bKSxUpdate1}), it is seen that $b_X(x_k^i) \sim \mathcal{N}({x_{i,k}|\hat x_{i,k}, P_{i,k}})$ is also Gaussian-distributed with its mean $\hat x_{i,k}$ and covariance $P_{i,k}$ being obtained by a nonlinear filter, such as extended Kalman filter~(EKF), that is,
\begin{equation}\label{eq:xP}
\hat x_{i,k} = \mathbb{E}\left(x_{i, k}|\textbf{y}_{i,k}, \textbf{h}_{i,k}, \textbf{R}_{i,k}\right), \quad P_{i,k} = \text{cov}\left(\hat{x}_{i, k},\hat{x}_{i,k}|\textbf{y}_{i,k}, \textbf{h}_{i,k}, \textbf{R}_{i,k}\right).
\end{equation}

The belief $b_X(x_{i,1:K})$ of $x_i$ for a time sequence $1:K$ is derived as
\begin{equation}\label{eq:bx1k}
b_X(x_{i, 1:K}) =  \prod_{k=1}^K b_X(x_{i,k}) = \prod_{k=1}^K \mathcal{N}\left(x_{i,k}|\hat x_{i, k|1:K}, P_{i,k|1:K} \right).
\end{equation}
with the mean $\hat x_{i, k|1:K}$ and the covariance $P_{i,k|1:K}$ being obtained by a nonlinear fixed-interval smoother, such as Extended Rauch-Tung-Striebel Smoother~(ERTSS).

\subsubsection{Derivation of Belief $b_E(E)$}
\begin{figure}[!htp]
\centering
\includegraphics[width=0.55\textwidth]{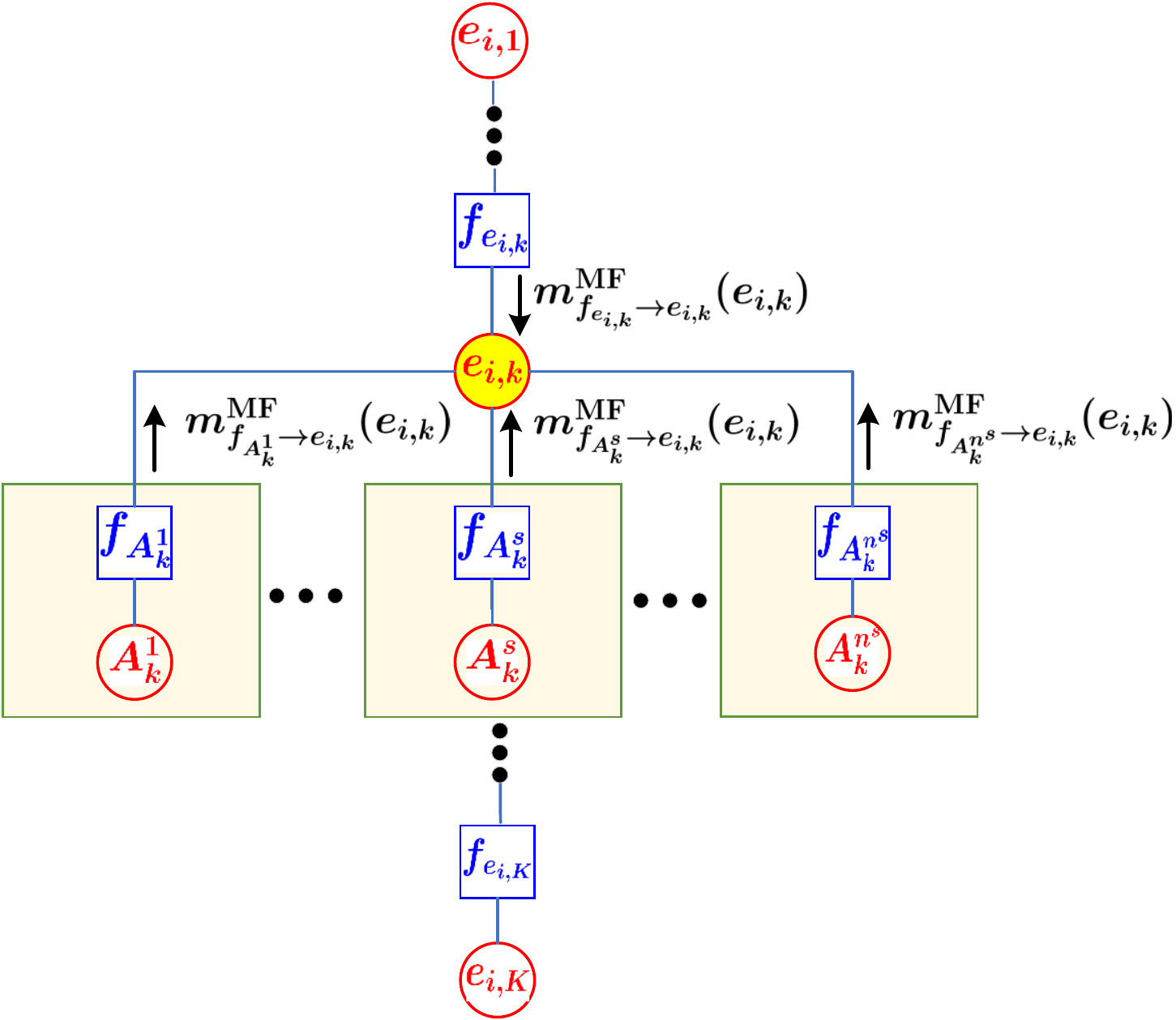}
\caption{The target visibility state estimation subgraph of $b_E(e_{i,k})$.}
\label{Fig3}
\end{figure}

Like the target kinematic state, each target appears and disappears independently, the belief of the joint visibility state of targets can be factorized as,
\begin{equation}\label{eq:bVS}
b_E(E) = \prod_{i = 1}^{n^x}b_E(e_{i,1:K}) = \prod_{i = 1}^{n^x} \prod_{k=1}^K b_E(e_{i,k}).
\end{equation}
The corresponding subgraph of the target visibility state estimation $b_E(e_{i,k})$ is shown in
Fig.~\ref{Fig3}, where the to-be-considered variable nodes are $e_{i,k}, i = 1, \ldots, n^x, k = 1, \ldots, K$.
For each variable node $e_{i,k}$, connect it with $n^s + 1$ factor nodes,
$\mathcal{S}(e_{i,k}) = \big\{f_{e_{i,k}}, f_{A_k^1}, \ldots, f_{A_k^{n^s}}\big\}$.
The sets of variable nodes connected to each factor node are $\mathcal{S}(f_{e_{i,k}}) = \left\{e_{i,k}, e_{i,k-1}\right\}$ and $\mathcal{S}(f_{A_k^1}) = \left\{e_{i,k}, A_k^s \right\}$, respectively.
According to the message-computation rules given in Eq.~(\ref{Eq-update}),
the messages from each factor nodes in $\mathcal{S}(e_{i,k})$ to the variable node $e_{i,k}$ are calculated as
\begin{align}
m^{\text{MF}}_{f_{e_{i,k}} \rightarrow e_{i,k}}(e_{i,k}) = & \exp\left(\sum \limits_{e_{i, k-1} = 0}^1 n_{e_{i,k-1} \rightarrow f_{e_{i,k}}}(e_{i,k-1}) \ln T_k \right) = T_k, \\
m^{\text{MF}}_{f_{A_k^s} \rightarrow e_{i,k}}(e_{i,k}) = & \exp \left( \sum_{\tau = 1}^{n^{m}} \sum_{a_{i,0,\tau,k}^s = 0}^1 n_{a_{i,0,\tau,k}^s \rightarrow f_{A_k^s}}(a_{i,0,\tau,k}^s) \ln p(A_k^s|e_{i,k}) \right) \\ \nonumber
\propto& \exp \left(\sum_{\tau = 1}^{n^{m}}\Bigg(\left(1 - \hat a_{i,0,\tau,k}^s\right) \ln\left(P_d^{\tau,s}(e_{i,k})\right) + \hat a_{i,0,\tau,k}^s \ln\left(1 - P_d^{\tau,s}(e_{i,k})\right) \Bigg)\right).
\end{align}
According to Eq.~(\ref{eq:bx}), the belief $b_E(e_k^i)$ can be computed as
\begin{align}
b_E(e_{i,k}) \propto & m^{\text{MF}}_{f_{e_{i,k}} \rightarrow e_{i,k}}(e_{i,k}) \times \prod_{s=1}^{n^s} m^{\text{MF}}_{f_{A_k^s} \rightarrow e_{i,k}}(e_{i,k}) \\ \nonumber
    = & T_k \underbrace{\exp \left(\sum_{s = 1}^{n^s}\sum_{\tau = 1}^{n^{m}}\Bigg(\left(1 - \hat a_{i,0,\tau,k}^s\right) \ln\left(P_d^{\tau,s}(e_{i,k})\right) + \hat a_{i,0,\tau,k}^s \ln\left(1 - P_d^{\tau,s}(e_{i,k})\right) \Bigg)\right)}_{\xi_{k}(e_{i,k})}.
\end{align}

The belief $b_E(e_{i, 1:K})$ of target visibility state for a time sequence $1:K$, is derived as
\begin{equation}\label{eq:bee1k}
b_E(e_{1:K}^i) = \prod_{k = 1}^K b_E(e_{i,k}) = \pi_{e_{i,1}} \xi_{1}(e_{i,1}) \prod_{k = 2}^K T_k \xi_{k}(e_{i,k}).
\end{equation}
It is seen that the belief $b_E(e_{i,1:K})$ follows an HMM with the indirect observation sequence $\left\{\xi_{1}(e_{i,1}), \ldots, \xi_{K}(e_{i,K})\right\}$, and the estimation of $b_E(e_{i, 1:K})$ can be soloved by a forward-backward algorithm~\cite{rabiner1989tutorial}.
The track management decisions, including track confirmation, maintenance and termination, can be made by comparing the probability of visibility state $b_E(e_{k}^i = 1)$ with different thresholds~\cite{li2001tracker}.

Remark 1. The performance of target detection can be improved due to the following reasons.
The indirect observation sequence $\xi_{k}(e_{i,k})$ integrates two kinds of information from all paths of all OTHRs,
i.e., $p_d^{\tau, s}(e_{i,k})$ and $a_{i,0, \tau,k}^{s}$, $\tau = 1, \ldots, n^m, s = 1, \ldots, n^s$.
The target-specific detection probability $p_d^{\tau, s}(e_{i,k})$, which depends on the target visibility state $e_{i,k}$, is used in this paper.
Compared with the constant detection probability $p_d^{\tau}$ which is prior information provided by the signal processing module, $p_d^{\tau, s}(e_{i,k})$ is posterior information considering the current information on target visibility state, which is benefit to improve the performance of target detection.
The association event $a_{i,0, \tau, k}^{s}$ contains the information that if target $i$ is visible or not for propagation path $\tau$ via OTHR $s$, which can be regarded as a prior information provided by measurements.

\subsubsection{Derivation of Belief $b_U(U)$}
\begin{figure}[!htp]
\centering
\includegraphics[width=0.25\textwidth]{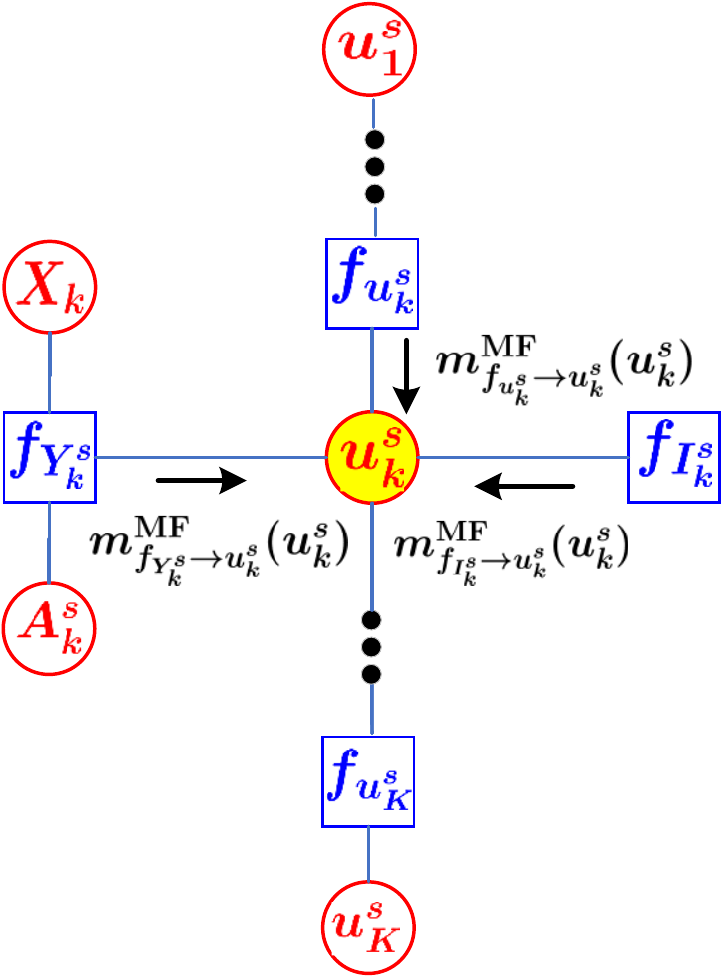}
\caption{The local ionospheric height identification subgraph of $b_U(u_k^s)$.}
\label{Fig4}
\end{figure}

Based on the assumption that the local ionospheric height is radar-specific, the belief of local ionospheric height can be factorized as
\begin{equation}
b_U(U) = \prod_{s = 1}^{n^s}b_U(u^s_{1:K}) = \prod_{s=1}^{n^s}\prod_{k=1}^K b_U(u_k^s),
\end{equation}
and the local ionospheric height subgraph that corresponds to the belief $b_U(u_k^s)$ is shown in Fig.~\ref{Fig4}.
The to-be-considered variable nodes of the local ionospheric height subgraph are $u_k^s, s = 1, \ldots, n^s, k = 1, \ldots, K$.
For each variable $u_k^s$, connect it with three factor nodes, $\mathcal{S}(u_k^s) = \left\{f_{u_k^s}, f_{Y_k^s}, f_{I_k^s}\right\}$.
The sets of variable nodes connected each factor node are $\mathcal{S}\left(f_{u_k^s}\right) = \left\{u_k^s, u_{k-1}^s \right\}$, $\mathcal{S}\left(f_{Y_k^s}\right) = \left\{X_k, A_k^s, u_k^s\right\}$, and $\mathcal{S}\left(f_{I_k^s}\right)  = \left\{u_k^s\right\}$, respectively. According to Eq.~(\ref{Eq-update}),
the messages from each factor nodes in $\mathcal{S}(u_k^s)$ to the variable node $u_k^s$ are given as
\begin{align}
m^{\text{MF}}_{f_{u_k^s} \rightarrow u_k^s}(u_k^s) =& \exp\left(\int_{u_{k-1}^s}  n_{u_{k-1}^s \rightarrow f_{u_k^s}}\left(u_{k-1}^s\right)\ln p\left(u_k^s|u_{k-1}^s\right) d_{u_{k-1}^s} \right) \label{eq:F2VU1}, \\
m^{\text{MF}}_{f_{Y_k^s} \rightarrow u_k^s}(u_{k}^s) =& \exp \Big(\sum_{i = 1}^{n^x}\sum_{j = 1}^{n_k^{e,s}}\sum_{\tau = 1}^{n^{m}} n_{a_{i,j,\tau,k}^s \rightarrow f_{Y_k^s}}\left(a_{i,j,\tau,k}^s\right) \\ \nonumber
&\times \int_{x_{i,k}} n_{x_{i,k} \rightarrow f_{Y_k^s}}\left(x_{i,k}\right) \ln p\left(y^s_{j,k}|x_{i,k}, u_k^s, a_{i,j,\tau,k}^s\Big) d_{x_{i,k}} \right) \label{eq:F2VU2}, \\
    m^{\text{MF}}_{f_{I_k^s} \rightarrow u_k^s}(u_k^s) =& p(I_k^s|u_k^s).
\end{align}
Noting that the variable-to-factor messages $n_{i \rightarrow a}(x_i) = b_i(x_i), \forall a \in \mathcal{F}_{\text{MF}} $ and $i \in \mathcal{S}(a)$,
we have
\begin{equation}\label{nu}
n_{u_{k-1}^s \rightarrow f_{u_k^s}}\left(u_{k-1}^s\right) = b_U(u_{k-1}^s), \quad
n_{x_{i,k} \rightarrow f_{Y_k^s}}\left(x_{i,k}\right) = b_{X}(x_{i,k}).
\end{equation}
Substituting Eq.~(\ref{nu}) into Eqs.~(\ref{eq:F2VU1}) and (\ref{eq:F2VU2}), yields
\begin{align}
m^{\text{MF}}_{f_{u_k^s} \rightarrow u_k^s}(u_k^s) =& \mathcal{N}\left(u_k^s|B_k \hat u_{k-1}^s, B_k \Sigma_{k-1}^s B_k^T + \mathcal{Q}_k \right), \\
m^{\text{MF}}_{f_{Y_k^s} \rightarrow u_k^s}(u_{k}^s) =& \prod_{i = 1}^{n^x}\prod_{j=1}^{n_k^{e,s}}\prod_{\tau=1}^{n^{m}}p\left(y_{j,k}^s|\hat x_{i,k}, u_{\tau,k}^{s}\right)^{\hat a_{i,j,\tau,k}^s} \label{eq:F2VU2Final},\\
m^{\text{MF}}_{f_{I_k^s} \rightarrow u_k^s}(u_k^s) =& \mathcal{N}\left(I_k^s|C_{k}^s u_k^s, W_k^s \right),
\end{align}
where $\hat u_{k-1}^s =\langle u_{k-1}^s \rangle_{b_U(u_{k-1}^s)}$ and $\hat x_{i, k} =\langle x_{i,k} \rangle_{b_X(x_{i,k})}$ are the expectation of $u_{k-1}^s$ and $x_{i,k}$ taken over corresponding beliefs.

Similar to Eq.~(\ref{Mlinear}), the nonlinear measurement function $p\left(y_{j,k}^s|\hat x_{i,k}, u_{\tau,k}^{s}\right)$ in Eq.~(\ref{eq:F2VU2Final}) is approximated by Gaussian PDF, i.e.,
\begin{equation}\label{MLinearU}
p\left(y_{j,k}^s|\hat x_{i,k}, u_{\tau,k}^{s}\right) \approx \mathcal{N}\left(y_{j,k}^s|h_k\left(\hat x_{i,k}, l^s, u_{\tau,k}^{s}\right), R_{\tau, x, k}^{s}  \right),
\end{equation}
where $ R_{\tau, x, k}^{s} = R_{\tau,k}^{s} + R_{x, k}^{s}$, and $R_{x,k}^{s} = J_x^{\tau,s} P_k (J_x^{\tau,s})^T$ with $J_x^{\tau,s}$ being given by Eq.~(\ref{eq:JXTau}).

According to Eq.~(\ref{eq:bx}), the belief $b_U(u_k^s)$ can be computed by multiplying all the incoming factor-to-variables messages as follows,
\begin{equation}
\begin{split} \label{BU}
b_U(u_k^s) \propto &  m^{\text{MF}}_{f_{u_k^s} \rightarrow u_k^s}(u_k^s) \times m^{\text{MF}}_{f_{Y_k^s} \rightarrow u_k^s}(u_{k}^s) \times m^{\text{MF}}_{f_{I_k^s} \rightarrow u_k^s}(u_k^s) \\
\approx & \mathcal{N}\left(u_k^s|B_k \hat u_{k-1}^s, B_k \Sigma_{k-1}^s B_k^T + \mathcal{Q}_k \right) \mathcal{N}\left(I_k^s|C_{k}^s u_k^s, W_k^s \right) \\ \nonumber
&\times \prod_{i = 1}^{n^x}\prod_{\tau=1}^{n^{m}}\mathcal{N}\left(\bar y_{i,\tau, k}^{s}|h_k\left(\hat x_{i,k}, l^s, u_{\tau,k}^{s} \right), \bar R_{i,\tau, x, k}^{s}  \right),
\end{split}
\end{equation}
where the synthetic measurement $\bar y_{i,\tau, k}^{s}$ is given by Eq.~(\ref{App-bar-y}) and the corresponding covariance $\bar R_{i,\tau, x, k}^{s}$ is defined as
\begin{equation}
\bar R_{i,\tau, x, k}^{s} = \dfrac{R_{\tau, x, k}^{s}}{1 - \hat a_{i,0,\tau,k}^{s}}.
\end{equation}
From Eq.~(\ref{BU}), it is seen that $b_U(u_k^s) \sim \mathcal{N}(u_k^s|\hat u_k^s, \Sigma_k^s)$ is Gaussian distributed with its mean and covariance being obtained by a nonlinear filter, such as UKF. That is,
\begin{equation}\label{Mean-U}
\begin{split}
\hat u_k^s =& \mathbb{E}\left(u_k^s|I_k^s, \bar y_{1, 1, k}^{s}, \ldots, y_{n^x, n^m, k}^{s} \right), \\
\Sigma_k^s =& \text{cov}\left(\hat u_k^s, \hat u_k^s|I_k^s, \bar y_{1,1,  k}^{s}, \ldots, y_{n^x, n^m, k}^{s}\right).
\end{split}
\end{equation}

The belief $b_U(u_{1:K}^s)$ of ionospheric heights state for a time sequence $1:K$, is derived as
\begin{equation}\label{eq::BuFinal}
b_U(u_{1:K}^s) = \prod_{k=1}^K b_U(u_k^s) = \prod_{k=1}^{K} \mathcal{N}\left(u_k^s|\hat u_{k|1:K}^s, \Sigma_{k|1:K}^s\right).
\end{equation}
In this case, the nonlinear filter in Eq.~(\ref{Mean-U}) can be replaced with a nonlinear fixed-interval smoother, such as ERTSS can be exploited.
The performance of ionospheric height identification can be improved by combining two kinds of information.
One is the direct measurement of ionosondes $I_k^s$, and the other is the indirect information from primary OTHR whereas the estimated target kinematic state and its corresponding measurements $\left\{\hat x_{i,k}, \bar y_{i, \tau, k}^{s}\right\}$, $i = 1, \ldots, n^x, \tau = 1, \ldots, n^{m},$ are feedback to identify the ionospheric height.

\subsubsection{Derivation of Belief $b_A(A)$}
We assume that the local multipath data association is independent over different OTHRs and different scans.
Accordingly, the belief on multipath data association $b_A(A)$ is factorized as,
\begin{equation}
b_A(A) = \prod_{s = 1}^{n^s}\prod_{k=1}^K b_A(a_{k}^s).
\end{equation}

\begin{figure}[!htp]
\centering
\includegraphics[width=0.55\textwidth]{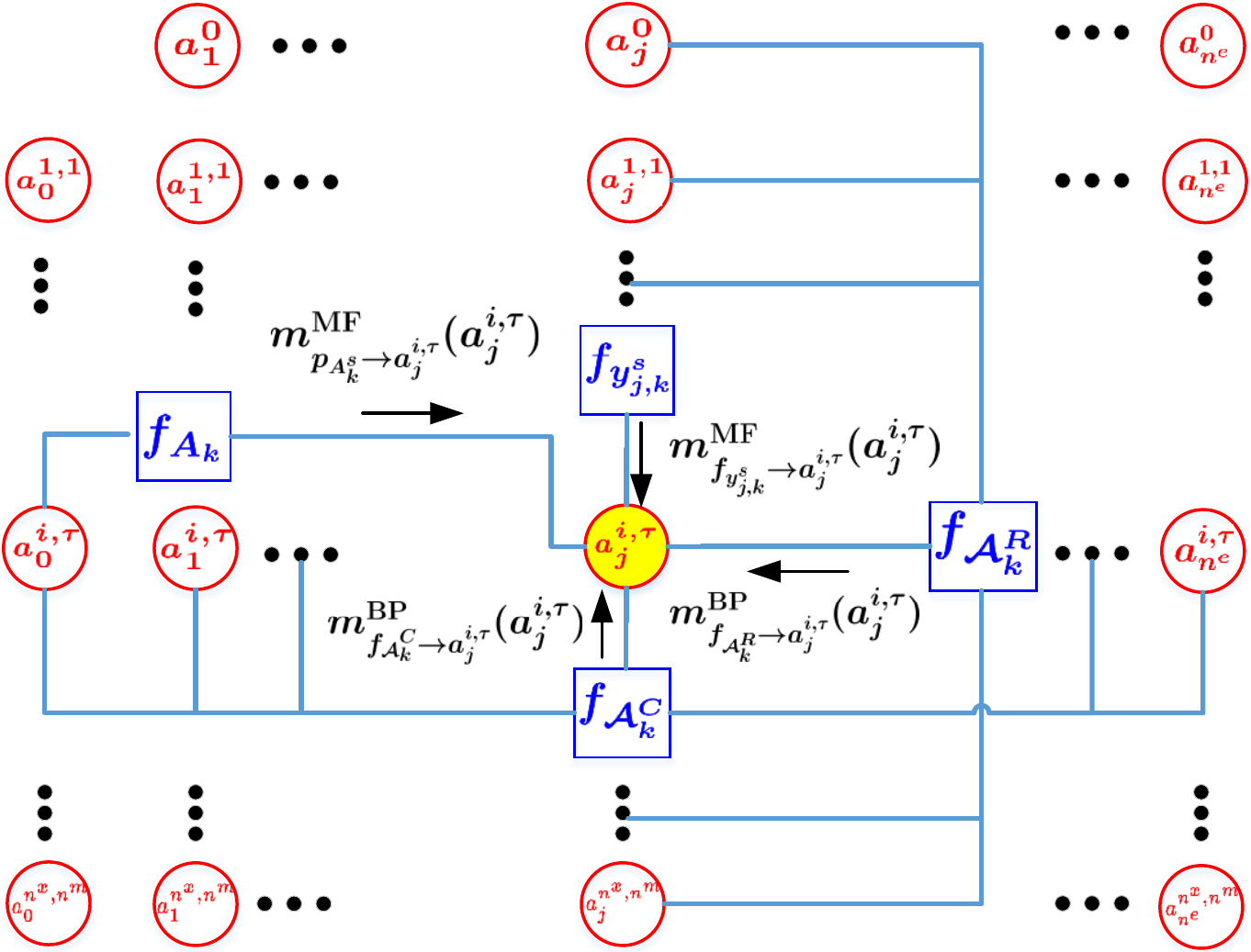}
\caption{The multipath data association subgraph of $b_A(a_{i,j,\tau,k}^s)$. The time index $k$ and OTHR index $s$ are omitted. The short notations are used: $a_j^{i,\tau} = a_{i,j,\tau,k}^s$ and $a_j^{0} = a_{0,j,k}^{s}$.}
\label{Fig5}
\end{figure}

Fig.~\ref{Fig5} shows the multipath data association subgraph corresponding to the belief $b_A(a_{k}^s)$, which consists of the variable nodes $a_{i,j,\tau,k}^s, i = 0, 1, \ldots, n^x, j = 0, \ldots, n_k^{e,s}, \tau = 1, \ldots, n^{m}, k = 1, \ldots, K$.
There are four factor nodes neighboring to variable node $a_{i,j,\tau,k}^{s}$, i.e., $\mathcal{S}\left(a_{i,j,\tau,k}^{s}\right) = \big \{f_{A_k^s}, f_{Y_k^s}, f_{\mathcal{A}_k^R}, f_{\mathcal{A}_k^C} \big \}$ where we denote $f_{\mathcal{A}_k^R} = \mathbb{I}\left(\sum_{i = 1}^{n^x}\sum_{\tau=1}^{n^{m}} a_{i,j,\tau,k}^{s} + a_{0, j, \tau, k}^s = 1\right)$ and $f_{\mathcal{A}_k^C} = \mathbb{I}\left(\sum_{j = 0}^{n_k^{e,s}} a_{i,j,\tau,k}^s = 1\right)$ for simplicity.
Note that $f_{\mathcal{A}_k} = f_{\mathcal{A}_k^R} \cap f_{\mathcal{A}_k^C}$.
The sets of variable nodes connecting to the corresponding factor node are $\mathcal{S}\left(f_{A_k}\right) = \{A_k, e_k\}$, $\mathcal{S}\left(f_{Y_k}\right) =  \left\{y_{j,k}^s, x_{i,k}, a_{i,j,\tau,k}^s  \right\}$, $\mathcal{S}\left(f_{\mathcal{A}_k^R}\right) = \left\{a_{i,j,\tau,k}^s \right\}_{i=1,\cdots,n^x, \tau = 1, \cdots, n^m} \cup \left\{a_{0,j,k}^s \right\}$ and $\mathcal{S}\left(f_{\mathcal{A}_k^C}\right) =  \left\{a_{i, j,\tau, k}^{s}\right\}_{j=0,\cdots,n_k^{e,s}}$, respectively.

By the message update rules given in Eq.~(\ref{Eq-update}), the messages that belong to the MF region ($a \in \mathcal{F}_{\text{MF}}$) can be calculated as
\begin{equation}
    \begin{split}
    m^{\text{MF}}_{f_{A_k^s} \rightarrow a_{i,j,\tau,k}^{s}}(a_{i,j,\tau,k}^{s}) =& \exp\left(\sum_{e_{i,k} = 0}^1 n_{e_{i,k} \rightarrow p_{A_k^s|e_{i,k}}} (e_{i,k}) \ln p(A_k^s|e_{i,k}) \right) \\ \nonumber
    =& \begin{cases}
    \left\langle (1 - P_d^{\tau, s}(e_{i,k}))^{a_{i,j,\tau,k}^s} \right\rangle_{b_E(e_{i,k})}, & \forall i > 0, j = 0, \tau > 0 \\
    \left\langle (P_d^{\tau, s}(e_{i,k}))^{a_{i,j,\tau,k}^s} \right\rangle_{b_E(e_{i,k})}, & \forall i > 0, j >0, \tau > 0
    \end{cases}
    \end{split}
\end{equation}
\begin{equation}
    \begin{split}
    m^{\text{MF}}_{f_{y_{j,k}^s} \rightarrow a_{i,j,\tau,k}^{s}}(a_{i,j,\tau,k}^{s}) =& \exp \Bigg(\int \int n_{x_{i,k} \rightarrow f_{y_{j,k}^s}}(x_{i,k})n_{u_k^s \rightarrow f_{y_{j,k}^s}}(u_k^s) \ln p(y_{j,k}^s|x_{i,k}, u_k^s, a_{i,j,\tau,k}^s) d_{x_{i,k}} d_{u_{k}^s} \Bigg)\\
    \propto & \exp \Bigg( a_{0, j,k}^{s} \ln (1 / V_k^s) + a_{i,j,\tau,k}^s\ln p\left(y_{j,k}^s|\hat x_{i,k}, \hat u_{\tau,k}^{s}\right) \Bigg) \\ \nonumber
    =& \begin{cases}
    {V_k^s}^{-a_{0,j,k}^{s}}, & \forall i = 0, j > 0 \\
    \exp\left(a_{i,j,\tau,k}^s\mathcal{X}^{s}_{i,j,\tau,k}\right),  & \forall i > 0, j>0, \tau > 0
    \end{cases}
    \end{split}
\end{equation}
with $\mathcal{X}^{s}_{i,j,\tau,k} = \ln p\left(y_{j,k}^s|\hat x_{i,k}, \hat u_{\tau,k}^{s}\right)$.
Similar to Eq.~(\ref{Mlinear}) and Eq.~(\ref{MLinearU}), due to the nonlinear function $h_k(\cdot)$, the PDF $p\left(y_{j,k}^s|\hat x_{i,k}, \hat u_{\tau,k}^{s}\right)$ is approximated by Gaussian PDF, i.e.,
\begin{equation}
p\left(y_{j,k}^s|\hat x_{i,k}, \hat u_{\tau,k}^{s}\right) \approx \mathcal{N}\left(y_{j,k}^s|h_k\left(\hat x_{i,k}, l^s, \hat u_{\tau,k}^{s}\right), R_{i,j,\tau,k}^{s}\right),
\end{equation}
where $R_{i,j,\tau,k}^{s} = R_{\tau,k}^{s} + R_{x,k}^{s} + R_{u,\tau,k}^{s}$.
Thus, the parameter $\mathcal{X}^{s}_{i,j,\tau,k}$ can be calculated as
\begin{equation}
\begin{split}
\mathcal{X}^{s}_{i,j,\tau,k} =&-\frac{1}{2} \text{Tr} \left\{\left(R_{i, j,\tau,k}^{s}\right)^{-1}\left(\left(y_{j,k}^s - h_k\left(\hat{x}_{i, k},l^s, \hat u_{\tau,k}^{s}\right)\right)\left(y_{j, k}^s - h_k\left(\hat{x}_{i,k}, l^s, \hat u_{\tau,k}^{s}\right)\right)^{\rm T}\right) \right\} \\
& + \frac{n_y}{2}\ln(2\pi) + \frac{1}{2}\ln \left|R_{i, j, \tau,k}^{s}\right|.
\end{split}
\end{equation}

For the messages belong to the BP region ($a \in \mathcal{F}_{\text{BP}}$), we have
\begin{equation}\label{eq:mBP}
\begin{split}
\mu^{\text{BP}}_{f_{\mathcal{A}_k^R} \rightarrow a_{i,j,\tau,k}^s}(a_{i,j,\tau,k}^s) = \sum_{a_{i,0,\tau,k}^s}\!\! \cdots \!\!\sum_{a_{i,j-1,\tau,k}^s} \sum_{a_{i,j+1,\tau,k}^s}\!\! \cdots \!\! \sum_{a_{i,n_k^{e,s},\tau,k}^s} f_{\mathcal{A}_k^R}
\prod_{{j_1 = 0(j_1 \neq j)}}^{n_k^{e,s}}n_{a_{i,j,\tau,k}^s \rightarrow f_{\mathcal{A}_k^R}}(a_{i,j,\tau,k}^s).
\end{split}
\end{equation}
Recall that from the frame constraint, for each OTHR $s$, target $i$ either produces a measurement $j$ through a particular path $\tau$ or is missed.
That is, if $a_{i,j,\tau,k}^s = 1$, then $a_{i,j_1,\tau,k}^{s} = 0, j_1 = 0, \ldots, j-1, j+1, \ldots, n_k^{e,s}$.
Eq.~(\ref{eq:mBP}) can be rewritten as
\begin{equation}
    \begin{split}
    \mu^{\text{BP}}_{f_{\mathcal{A}_k^R} \rightarrow a_{i,j,\tau,k}^s}(a_{i,j,\tau,k}^s)
    \!\!=\!\! \begin{bmatrix}
    \mu^{\text{BP}}_{f_{\mathcal{A}_k^R} \rightarrow a_{i,j,\tau,k}^s}(0)  \\
    \mu^{\text{BP}}_{f_{\mathcal{A}_k^R} \rightarrow a_{i,j,\tau,k}^s}(1)
    \end{bmatrix}
    \!\!=\!\!
    \begin{bmatrix}
    \sum\limits_{\substack{j_1 = 1\\(j_1 \neq j)}}^{n_k^{e,s}} n_{a_{i,j_1,\tau,k}^{s} \rightarrow f_{\mathcal{A}_k^R}}(1) \!\! \prod\limits_{\substack{j_2 = 1\\(j_2 \neq j_1, j)}}^{n_k^{e,s}}\!\!n_{a_{i,j_2,\tau,k}^{s} \rightarrow f_{\mathcal{A}_k^R}}(0)  \\
    \prod\limits_{\substack{j_1 = 1\\(j_1 \neq j)}}^{n_k^{e,s}}n_{a_{i,j_1,\tau,k}^{s} \rightarrow f_{\mathcal{A}_k^R}}(0)
    \end{bmatrix}.
    \end{split}
\end{equation}

In a similar way, the message $m^{\text{BP}}_{f_{\mathcal{A}_k^C} \rightarrow a_{i,j,\tau,k}^{s}}(a_{i,j,\tau,k}^{s})~(i >0, j >0,\tau >0)$ can be rewritten as
\begin{equation}
\begin{split}
m^{\text{BP}}_{f_{\mathcal{A}_k^C} \rightarrow a_{i,j,\tau,k}^{s}}(a_{i,j,\tau,k}^{s})
\!\!=\!\!&\begin{bmatrix}
\sum \limits_{\substack{i_1 = 1, \tau_1 = 1 \\( i_1 \neq i, \tau_1 \neq \tau)}}^{n^x, n^{m}} n_{a_{i_1,j,\tau_1,k}^{s} \rightarrow f_{\mathcal{A}_k^R}}(1) \!\! \prod  \limits_{\substack{i_2 = 1, \tau_2 = 1\\( i_2 \neq i, \tau_2 \neq \tau \text{and} i_2 \neq i_1, \tau_2 \neq \tau_1)}}^{n^x, n^{m}}\!\! n_{a_{i_2,j,\tau_2,k}^{s} \rightarrow f_{\mathcal{A}_k^R}}(0) \\
    \prod \limits_{{i_1 \neq i, \tau_1 \neq \tau}}n_{a_{i_1,j,\tau_1,k}^{s} \rightarrow f_{\mathcal{A}_k^R}}(0)
\end{bmatrix}.
\end{split}
\end{equation}

According to Eq.~(\ref{eq:bx}), the belief $b_A\left(a_{i,j,\tau,k}^s\right)$ can be computed as
\begin{equation}\label{eq:baakij}
\begin{split}
b_A\left(a_{i,j,\tau,k}^s\right) \propto& m^{\text{MF}}_{f_{A_k^s} \rightarrow a_{i,j,\tau,k}^s}\left(a_{i,j,\tau,k}^s\right) \times m^{\text{MF}}_{f_{y_{j,k}^s} \rightarrow a_{i,j,\tau,k}^s}\left(a_{i,j,\tau,k}^s\right) \\
&\times m^{\text{BP}}_{f_{\mathcal{A}_k^R} \rightarrow a_{i,j,\tau,k}^{s}}\left(a_{i,j,\tau,k}^{s}\right) \times  m^{\text{BP}}_{f_{\mathcal{A}_k^C} \rightarrow a_{i,j,\tau,k}^{s}}\left(a_{i,j,\tau,k}^{s}\right).
\end{split}
\end{equation}

Accordingly, the expectation $\hat a_{i,j,\tau,k}^s$ is given by
\begin{equation}\label{eq:eakij}
\begin{split}
\hat a_{i,j,\tau,k}^s =& \dfrac{b_A(a_{i,j,\tau,k}^s = 1)}{b_A(a_{i,j,\tau,k}^s = 1) + b_A(a_{i,j,\tau,k}^s = 0)} \\
= & \dfrac{1}{1 + \dfrac{ m^{\text{MF}}_{f_{A_k^s} \rightarrow a_{i,j,\tau,k}^s}(0)}{m^{\text{MF}}_{f_{A_k^s} \rightarrow a_{i,j,\tau,k}^s}(1)}
    \times \dfrac{m^{\text{MF}}_{f_{y_{j,k}^s} \rightarrow a_{i,j,\tau,k}^s}(0)}{m^{\text{MF}}_{f_{y_{j,k}^s} \rightarrow a_{i,j,\tau,k}^s}(1)}
    \times \dfrac{m^{\text{BP}}_{f_{\mathcal{A}_k^R} \rightarrow a_{i,j,\tau,k}^s}(0)}{m^{\text{BP}}_{f_{\mathcal{A}_k^R} \rightarrow a_{i,j,\tau,k}^s}(1)}
    \times \dfrac{m^{\text{BP}}_{f_{\mathcal{A}_k^C} \rightarrow a_{i,j,\tau,k}^s}(0)}{m^{\text{BP}}_{f_{\mathcal{A}_k^C} \rightarrow a_{i,j,\tau,k}^s}(1)}}\\
    = & \dfrac{1}{1 + \exp\left(\!- \!\ln \bar m^{\text{MF}}_{f_{A_k^s} \rightarrow a_{i,j,\tau,k}^s} \! - \! \ln \bar m^{\text{MF}}_{f_{y_{j,k}^s} \rightarrow a_{i,j,\tau,k}^s} \!- \! \ln \bar m^{\text{BP}}_{f_{\mathcal{A}_k^R} \rightarrow a_{i,j,\tau,k}^s} \!- \! \ln \bar m^{\text{BP}}_{f_{\mathcal{A}_k^C} \rightarrow a_{i,j,\tau,k}^s}\right)}
\end{split}
\end{equation}
with
\begin{align}
    &\bar m^{\text{MF}}_{f_{A_k^s} \rightarrow a_{i,j,\tau,k}^{s}} = \dfrac{ m^{\text{MF}}_{f_{A_k^s}\rightarrow a_{i,j,\tau,k}^{s}}(1)}{ m^{\text{MF}}_{f_{A_k^s}\rightarrow a_{i,j,\tau,k}^{s}}(0)} = m^{\text{MF}}_{f_{A_k^s} \rightarrow a_{i,j,\tau,k}^{s}}(1),\label{eq:mMF1} \\
    &\bar m^{\text{MF}}_{f_{y_{j,k}^s}\rightarrow a_{i,j,\tau,k}^{s}} = \dfrac{ m^{\text{MF}}_{f_{y_{j,k}^s}\rightarrow a_{i,j,\tau,k}^{s}}(1)}{ m^{\text{MF}}_{f_{y_{j,k}^s}\rightarrow a_{i,j,\tau,k}^{s}}(0)} = m^{\text{MF}}_{f_{y_{j,k}^s}\rightarrow a_{i,j,\tau,k}^{s}}(1),\label{eq:mMF2} \\
    &\bar m^{\text{BP}}_{f_{\mathcal{A}_k^R} \rightarrow a_{i,j,\tau,k}^{s}} = \dfrac{ m^{\text{BP}}_{f_{\mathcal{A}_k^R} \rightarrow a_{i,j,\tau,k}^{s}}(1)}{ m^{\text{BP}}_{f_{\mathcal{A}_k^R} \rightarrow a_{i,j,\tau,k}^{s}}(0)}
    = \dfrac{1}{\sum\limits_{\substack{j_1 = 0 \\ (j_1 \neq j)}}^{n_k^{e,s}} n_{a_{i,j_1,\tau,k}^{s} \rightarrow
f_{\mathcal{A}_k^R}}(1)/n_{a_{i,j_1,\tau,k}^{s} \rightarrow f_{\mathcal{A}_k^R}}(0)}, \label{eq:mBP1} \\
    &\bar m^{\text{BP}}_{f_{\mathcal{A}_k^C} \rightarrow a_{i,j,\tau,k}^{s}} = \dfrac{ m^{\text{BP}}_{f_{\mathcal{A}_k^C} \rightarrow a_{i,j,\tau,k}^{s}}(1)}{ m^{\text{BP}}_{f_{\mathcal{A}_k^C} \rightarrow a_{i,j,\tau,k}^{s}}(0)}\label{eq:mBP2}
    = \dfrac{1}{\sum \limits_{\substack{i_1 = 0, \tau_1 = 1, \\ (i_1 \neq i, \tau_1 \neq \tau)}}^{n_k^x, n^{m}}
        n_{a_{i_1,j,\tau_1,k}^{s} \rightarrow f_{\mathcal{A}_k^C}}(1)/n_{a_{i_1,j,\tau_1,k}^{s} \rightarrow f_{\mathcal{A}_k^C}}(0)}.
\end{align}
According to the message update rules in Eq.~(\ref{Eq-update}), $n_{a_{i,j,\tau,k}^s \rightarrow f_{\mathcal{A}_k^R}}(a_{i,j,\tau,k}^s)$ and $n_{a_{i,j,\tau,k}^s \rightarrow f_{\mathcal{A}_k^C}}(a_{i,j,\tau,k}^s)$
in Eqs.~(\ref{eq:mBP1}), (\ref{eq:mBP2}) are
\begin{equation}\label{eq:na2fiR}
n_{a_{i,j,\tau,k}^s \rightarrow f_{\mathcal{A}_k^R}}(a_{i,j,\tau,k}^s) = \begin{cases}
    m^{\text{MF}}_{f_{A_k^s} \rightarrow a_{i,j,\tau,k}^s}(a_{i,0,\tau,k}^s), \quad \forall i > 0, j = 0, \tau > 0 \\
    m^{\text{MF}}_{f_{A_k^s} \rightarrow a_{i,j,\tau,k}^s}(a_{i,j,\tau,k}^s) m^{\text{MF}}_{f_{y_{j,k}^s} \rightarrow a_{i,j,\tau,k}^s}(a_{i,j,\tau,k}^s) m^{\text{BP}}_{f_{\mathcal{A}_k^C} \rightarrow a_{i,j,\tau,k}^{s}}(a_{i,j,\tau,k}^{s}),
    \\ \qquad \qquad  \qquad \qquad \qquad \forall i > 0, j > 0, \tau > 0
\end{cases}
\end{equation}
and
\begin{equation}\label{eq:na2fjC}
n_{a_{i,j,\tau,k}^s \rightarrow f_{\mathcal{A}_k^C}}(a_{i,j,\tau,k}^s) = \begin{cases}
    m^{\text{MF}}_{f_{y_{j,k}^s} \rightarrow a_{0,j,k}^s}(a_{0, j,k}^{s}),  \quad \forall i = 0, j > 0 \\
    m^{\text{MF}}_{f_{A_k^s} \rightarrow a_{i,j,\tau,k}^s}(a_{i,j,\tau,k}^s) m^{\text{MF}}_{f_{y_{j,k}^s} \rightarrow a_{i,j,\tau,k}^s}(a_{i,j,\tau,k}^s) m^{\text{BP}}_{f_{\mathcal{A}_k^R} \rightarrow a_{i,j,\tau,k}^{s}}(a_{i,j,\tau,k}^{s}), \\ \qquad \qquad  \qquad \qquad \quad \forall i > 0, j > 0, \tau > 0
\end{cases}
\end{equation}

Substituting Eqs.~(\ref{eq:na2fiR}), (\ref{eq:na2fjC}) into Eqs.~(\ref{eq:mBP1}), (\ref{eq:mBP2}), yields,
\begin{align}
&\bar m^{\text{BP}}_{f_{\mathcal{A}_k^R} \rightarrow a_{i,j,\tau,k}^{s}}(a_{i,j,\tau,k}^{s}) = \dfrac{1}{\bar m^{\text{MF}}_{f_{A_k^s} \rightarrow a_{i,0,\tau,k}^s} + \sum \limits_{\substack{j_1 =1\\(j_1 \neq j)}}^{n_k^{e,s}} \bar m^{ \text{MF}}_{f_{A_k^s} \rightarrow a_{i,j_1,\tau,k}^{s}}\bar m^{\text{MF}}_{f_{y_{j,k}^s} \rightarrow a_{i,j_1,\tau,k}^{s}} \bar m^{\text{BP}}_{f_{\mathcal{A}_k^C} \rightarrow a_{i,j_1,\tau,k}^{s}}} \label{eq:mfrakij}, \\
&\bar m^{\text{BP}}_{f_{\mathcal{A}_k^C} \rightarrow a_{i,j,\tau,k}^{s}}(a_{i,j,\tau,k}^{s}) \!=\! \dfrac{1}{\bar m^{\text{MF}}_{f_{y_{j,k}^s} \rightarrow a_{0,j,k}^s} \!\!\! + \!\!\! \sum \limits_{\substack{i_1=1, \tau_1=1 \\ (i_1 \neq i, \tau_1 \neq \tau)}}^{n^x, n^{m}} \bar m^{\text{MF}}_{f_{A_k^s} \rightarrow a_{i_1,j,\tau_1,k}^{s}} \bar m^{\text{MF}}_{f_{y_{j,k}^s} \rightarrow a_{i_1,j,\tau_1,k}^{s}} \bar m^{\text{BP}}_{f_{\mathcal{A}_k^R} \rightarrow a_{i_1,j,\tau_1,k}^{s}}} \label{eq:mfcakij}.
\end{align}

\subsection{Summary}
The proposed MP-OTHRs algorithm is summarized as \textbf{Algorithm}~\ref{alg}.
To achieve a trade-off between accuracy and latency, MP-OTHRs works in an online fashion using a sliding window.
\begin{algorithm}
\caption{The proposed MP-OTHRs algorithm}
\begin{algorithmic}[1]
\REQUIRE Measurements $Y_{k - \ell + 1 : k}, I_{k - \ell + 1 : k},  k \geq \ell$ with $\ell > 0$ being the window length;
\ENSURE Beliefs $b_X(X_{k-\ell+1:k})$, $b_E(E_{k-\ell+1:k})$, $b_A(A_{k-\ell+1:k})$, $b_U(U_{k-\ell+1:k})$;
\STATE \underline{\textbf{Initialization:}} initialize beliefs $b_i^{(0)}(x_i)$ for all $i \in \mathcal{I}_{\text{MF}} \backslash \mathcal{I}_{\text{BP}}$, i.e., $b_X^{(0)}(X_{k-\ell+1:k})$, $b_E^{(0)}(E_{k-\ell+1:k})$, $b_U^{(0)}(U_{k-\ell+1:k})$, and the maximum number of potential tracks $n^x_{k-\ell+1:k}$ during the sliding window; send the corresponding messages $n_{i \rightarrow a}(x_i) = b_i^{(0)}(x_i)$ to all factor nodes $a \in \mathcal{S}_{\text{MF}}(i)$. Let $r_{\text{max}}$ be the maximum number of iterations.
\FOR{each iteration $\iota = 1 : \iota_{\text{max}}$  }
\STATE  \underline{\textbf{Multipath data association:}} Calculate $b_A(a^{s(\iota)}_{i,j,\tau, t})$ and  $\hat a_{i,j,\tau,t}^{s(\iota)}, s = 1, \ldots, n^s, t = k-\ell+1, \ldots, k,$ iteratively via Eqs.~(\ref{eq:baakij}), (\ref{eq:eakij}) with messages given by Eqs.~(\ref{eq:mfrakij}), (\ref{eq:mfcakij}).
\STATE  \underline{\textbf{Ionospheric height identification:}} Calculate $b_U(u^{s(\iota)}_{k-\ell+1:k}), s = 1, \ldots, n^s$, via Eq.~(\ref{eq::BuFinal}).
\STATE \underline{\textbf{Target detection}:} Calculate $b_E(e^{(\iota)}_{i,k-\ell+1:k}), i = 1, \ldots, n_{k-\ell+1:l}^x$, via Eq.~(\ref{eq:bee1k}).
\STATE  \underline{\textbf{Target tracking:}} Calculate $b_X(x^{(\iota)}_{i, k-\ell+1:k}),i = 1, \ldots, n_{k-\ell+1:l}^x$, via Eq.~(\ref{eq:bx1k}).
\STATE \underline{\textbf{Iteration termination rule:}} the iteration terminates if the difference of beliefs between two consecutive iterations is less than the iteration threshold $\delta_T$.
\ENDFOR
\STATE Go to the next sliding window.
\end{algorithmic}
\label{alg}
\end{algorithm}


Compared with MR-MPTF, which is an open-loop, recursive processing and track-level fusion algorithm,
MP-OTHRs is a closed-loop, batch processing and measurement-level fusion algorithm.
The pros of MP-OTHRs are given as follows:
(1) MP-OTHRs adopts the UTM coordinate system to model the target kinematic state, which is more accurate than that of MR-MPTF;
(2) MP-OTHRs reduces information loss since it uses the pseudo-measurements to update the global target kinematic state directly;
(3) MP-OTHRs implements the estimation of target kinematic state and viability state via smoothers using a batch of measurements,
while MR-MPTF adopts a filter;
(4) MP-OTHRs is a joint optimization solution, and the information is exchanged among processing modules of latent variables;
(5) MP-OTHRs is more computationally effective than MR-MPTF by the fact that MP-OTHRs adopts LBP while MPTF uses the multiple hypothesis strategy to deal with the most time-consuming association problem.
As a result, the performance of MP-OTHRs is superior to that of MR-MPTF in the aspects of both target detection and tracking.

\subsection{Initialization} \label{sec:initialization}
Initial beliefs $b_X^{(0)}(X_{1:l})$, $b_E^{(0)}(E_{1:l})$, $b_U^{(0)}(U_{1:l})$ and the maximum number of potential targets~(tracks) $n^x_{1:\ell}$ are required for MP-OTHRs. We propose a  multisensor multipath measurements clustering approach for fast initialization of potential tracks.
The initialization procedure for the first sliding window $k \in [1, \ell]$ is given as follows.
\begin{itemize}
\item At time $k = 1$, tentative tracks are established via multisensor multipath measurement clustering, which consists of local track initialization and global track fusion. 1) For each OTHR $s, s = 1, \ldots, n^s$, we coarsely group the measurements into different subsets $\mathcal{Y}_{i,k}^{s} = \{y_{1,k}, \ldots, y_{N_{i,k}^{s},k}\}, i = 1, \ldots, N_k^{s}$ with $N_k^s$ and $N_k^{i,s}$ being the number measurements and number of measurements in the $i$th subset, respectively. To reduce the number of subsets, we assume that the number of measurements satisfies $1 < N_k^{i,s} \leq n^{m}$ since the maximum number of measurements from the same target is $n^{\tau, s}$ when a target is detected by all propagation paths. In each subset $\mathcal{Y}_{i,k}^{s}$, any two measurements are within a preset threshold vector $\rho^s$ conditioned on the assumption that they are from the same target via different paths. Each subset $\mathcal{Y}_{i,k}^{s}$ with at least two measurements are then utilized to initialize a local tentative track $x_{i,k}^{s}$. That is, for each subset $\mathcal{Y}_{i,k}^{s}$, there are total $n^m! / (n^m - N_k^{i,s})!$ measurement-path association hypothesis,
    and we transform the measurements in $\mathcal{Y}_{i,k}^{s}$ from radar slant coordinate system to UTM coordinate system by traversing all measurement-path association hypothesis, obtaining a set of the transformed path-dependent kinematic states.
    The path-dependent kinematic state that has the minimum average Mahalanobis distance is then selected and fused to obtain
    the local kinematic state estimate $\hat x_{i,k}^s$.
    The corresponding state covariance $P_{i,k}^s$ is pre-determined based on measurement noise covariance. Set the initial local target visibility probability $p(e_{i,k}^s  = 1) = N_k^{i,s} / n^{m}$.
    2) The global tracks are obtained by carrying out multisensor track association and fusion. That is, for any pair of local tracks $\{\hat x_{i,k}^{s_1}, P_{i,k}^{s_1}\}$ and $\{\hat x_{j,k}^{s_2}, P_{j,k}^{s_2}\}$ from different OTHR $s_1$ and $s_2$, the local tracks $i$ of OTHR $s_1$ and $j$ of OTHR $s_2$ are possibly from the same target if the Mahalanobis distance $d_{i,j} = (\hat x_{i,k}^{s_1} - \hat x_{j,k}^{s_2})^T(P_{i,k}^{s_1} + P_{j,k}^{s_2})^{-1}(\hat x_{i,k}^{s_1} - \hat x_{j,k}^{s_2})$ is less than the association threshold $\gamma$. To this end, an assignment matrix $D$ is constructed by assigning $D_{i,j} = 1$ if $d_{i,j} < \gamma$, $D_{i,j} = 0$ otherwise.
    The corresponding track-to-track association problem with $D$ is then solved by S-D assignment techniques~\cite{Bar1995Multitarget}.
    The global target tracks with kinematic states $\{\hat x_{i,k}, P_{i,k}\}$s consist of the fused tracks from those associated local tracks and
    the unassociated local tracks.
    Meanwhile, the global initial target visibility state $p(e_{i, k} = 1) = \sum_{s = 1}^{n^s} p(e^s_{i, k} = 1) / n^s$. The initial ionospheric height $\hat u_k^s = I_k^s$.
\item Starting from $k = 2$, for each tentative track $i$, transform the global kinematic state from UTM coordinate system to radar slant coordinate system, and select candidate local multipath measurements individually using gating technique, establish the pseudo-measurement via LBP, and update the kinematic state $\{\hat x_{i, k}, P_{i,k}\}$ by a filter. Meanwhile, the visibility probability $p(e_{i, k})$ and ionospheric height $\hat u_k^s$ are recursively updated by using forward algorithm and Kalman filter, respectively. The measurements that do not fall into the validation gates of any tracks are used to initialize new tracks.
\item For the batch window $[1, \ell]$, manage tracks based on $p(E_{1:\ell})$. Specifically, if the average visibility probability of target $i$ larger than the threshold $\delta_c$, track $i$ is confirmed; or less than the threshold $\delta_c$ in three successive scans, the track $i$ is deleted. $n_{1:\ell}^x$ is the total number of confirmed potential tracks over the batch window $[1, \ell]$.
    \end{itemize}


\section{Simulation and Analysis}\label{sec:simulation}
\subsection{Simulation Scenario}
\begin{figure}[!htbp]
    \centering
    \includegraphics[width=0.45\textwidth]{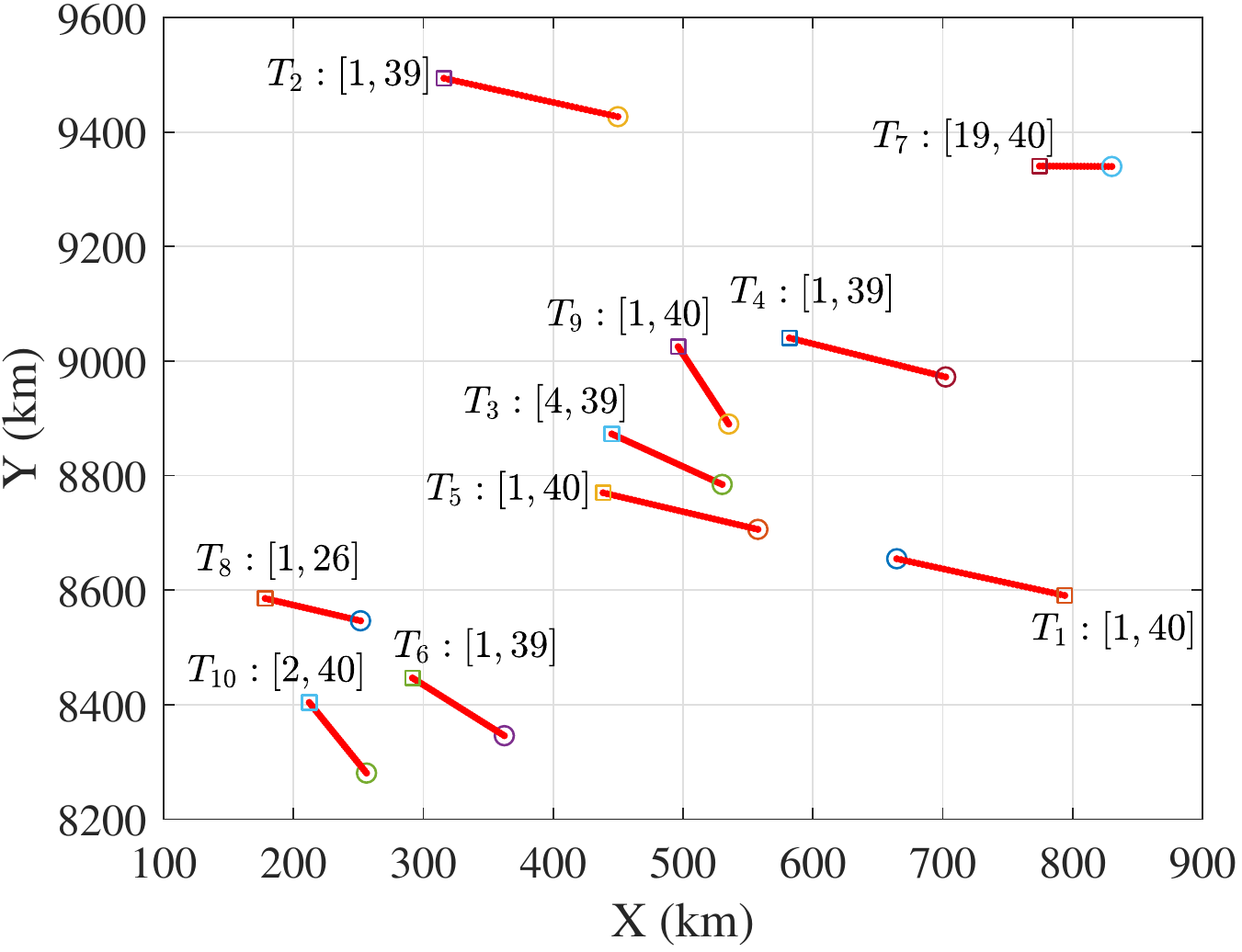}
    \caption{Target trajectories and lifetime in UTM coordinate system. $T_i : [k_1, k_2]$ represents that target $i$ appears at time $k_1$ and disappears at times $k_2$. $\circ$ and $\square$ represent the start point and the end point of a target, respectively. }\label{Trajectory}
\end{figure}

We consider the simulation scenario of multitarget tracking with an OTHR network consisting of two OTHRs.
Ten targets move in the overlapping region of the two OTHRs.
The detail information of target kinematic state and lifetimes is given in Fig.~\ref{Trajectory}.
MP-OTHRs is compared with MR-MPTF.
Note that MR-MPTF uses the real value of ionospheric height.
We also compare MP-OTHRs with a single OTHR measurement, referred as MP-OTHR1~(only using measurements from OTHR1) and MP-OTHR2~(only using measurements from OTHR2), respectively.

The detailed parameters setting are given as follows: the number of radars $n^s = 2$; two ionospheric layers~(E-layer and F-layer) and the number of propagation paths $n^{m} = 4$; the number of targets $n^x  = 10$; the radar receiver parameters  $l^1 = (143.20\degree, -24.29\degree, 325\degree, 100~\text{km})$ and $l^2 = (122.01\degree, -28.33\degree, 350\degree, 100~\text{km})$; the region sizes of range, azimuth and range rate are [1000, 3000]~km, [-0.3, 0.3]~rad, and [-0.3, 0.3]~km/s. For $s = 1, 2$, $\tau = 1, 2, 3, 4$,
the sampling period $T^s = 15 \text{s}$, the clutter density $\lambda^s = 1e-5$~(the expected clutter number is 21 per scan), the mean ionospheric height $h^s_E = 100$~km, $h^s_F = 260$~km, and $u_1^{s} = [h_E^s, h_E^s]$, $u_2^{s} = [h_E^s, h_F^s]$, $u_3^{s} = [h_F^s, h_E^s]$, $u_4^{s} = [h_F^s, h_F^s]$; the radar measurement noise covariance $R_{\tau}^{s} =  \text{diag}(25~\text{km}^2, 1e-6~\text{km}^2/\text{s}^2, 9e-6~\text{rad}^2)$, the ionospheric measurement noise covariance $W^s = \text{diag}(100~\text{km}^2, 100~\text{km}^2)$.
The state transition matrix
$F = I_2 \otimes \left[1, 15; 0, 1\right]$
and $B^s = I_2$, the ionospheric measurement matrix $C_k = I_2$, the process noise covariance  $Q =  \text{diag}(1e-6~\text{rad}^2, 1e-8~\text{rad/s}^2,1e-6~\text{rad}^2, 1e-8~\text{rad/s}^2)$ and $\mathcal{Q}^s = \text{diag}(1~\text{km}^2, 1~\text{km}^2)$, the initial state covariance $P_0 = \text{diag}(25~\text{km}^2, 0.04~\text{km/s}^2, 25~\text{km}^2, 0.04~\text{km/s}^2)$.
The MP iteration threshold $\delta_T = 1e-5$ and $r_{\text{max}} = 4$. The LBP iteration threshold $\delta_l = 1e-6$ and $\iota_{\text{max}} = 1000$. Tentative track initialization threshold $\rho^{s} = [80~\text{km}, 0.01~\text{km/s}, 0.1~\text{rad}]$. Track confirmation threshold $\delta_{c} = 0.9$. A track $i$ is confirmed if the target visibility state $p(e_{i} = 1) > \delta_{c}$ and deleted if $p(e_{i} = 1) < \delta_{c}$ over three successive scans.
The initial target visibility state probability $\pi$ is given by the initialization, and its transition probabilities $\text{Pr}(0|0) = \text{Pr}(1|1) = 0.85$, $\text{Pr}(0|1) = \text{Pr}(1|0) = 0.15$. The target visibility state-dependent detection probability $P_d^{\tau, s}(1) = p_d^{\tau,s}$ and $P_d^{\tau, s}(0) = 0.1$. The window length $\ell = 3$, the sliding window step is 1, and the gate probability $p_g = 0.971$.
MR-MPTF implements the single-path target tracking with global nearest neighbour~(GNN) tracker, and then carries out the track fusion using multiple hypothesis track fusion~(MPTF).
GNN uses 3/5 logic to initialize tracks and 3/3 logic to terminate tracks.
MPTF keeps first three best hypothesis.
The statistical performance comparison of MP-OTHRs and MR-MPTF are given with 100 Monte Carlo runs by varying detection probability.

To evaluate the performance of the four algorithms, it is required to find the association between targets and tracks.
We declare that a track is associated to a target if the average difference in both X position and Y position is less than 10~km.
Tracks with at least length ten are used to calculate the performance metrics that are given as follows.
(1) Number of True Tracks (NTT $\uparrow$): A track is detected as a true one if it is assigned to a target.
If more than one track are assigned to the same target, the maximum length track is selected and the rest are ignored.
(2) Track Probability of Detection (TPD $\uparrow$): Ratio of the length of a true track to the lifetime of its corresponding target, which indicates the stable tracking capability of an algorithm.
(3) Number of False Tracks (NFT $\downarrow$): A track is detected as a false one if it is not assigned to any target. The false tracks consist of the tracks arising from clutter and the multipath tracks that are not assigned to any target.
As stated in~\cite{sathyan2013multiple}, misleading path association can give rise to multipath tracks that are far away from the true trajectory.
(4) Confirmed Track Latency (CTL $\downarrow$): the time delay of the confirmation of a true track.
A good target detection capability is achieved by an algorithm if the value of CTL is small.
(5) Average Euclidean Error of Target Position~(AEEP $\downarrow$): 
(6) Average Euclidean Error of Target Speed~(AEES $\downarrow$).
(7) Average Euclidean Error of Ionospheric Height (AEEH $\downarrow$).
(8) Mean Optimal Subpattern Assignment (MOSPA $\downarrow$)~\cite{Schuhmacher2008A}:
(9) Total Execution Time (TET $\downarrow$).
For the more detailed definitions of TPD, NFT, AEE, CTL, and TET, refer to \cite{Gorji2011Performance}.
The metrics TPD, CTL, AEEP, AEES, and MOSPA are averaged overall targets, and all of the metrics are averaged over Monte Carlo runs.
The notation $\uparrow$~($\downarrow$) indicates the higher~(lower) value the metric, the better~(worse) the performance is.

\begin{figure}[!htbp]
    \centering
    \subfloat[Radar 1]{\label{fig12-a}\includegraphics[scale=0.55]{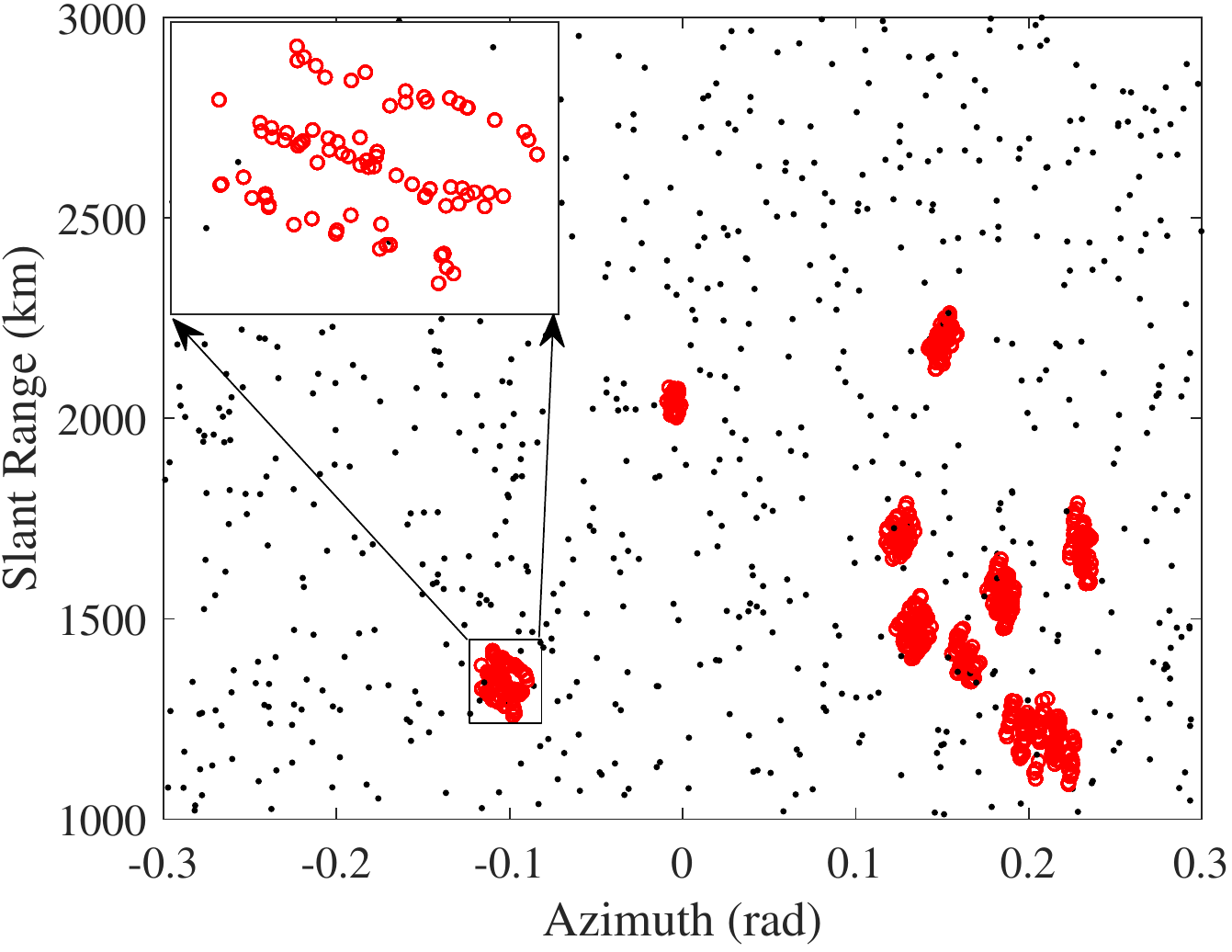}} \hspace{10pt}
    \subfloat[Radar 2]{\label{fig12-b}\includegraphics[scale=0.55]{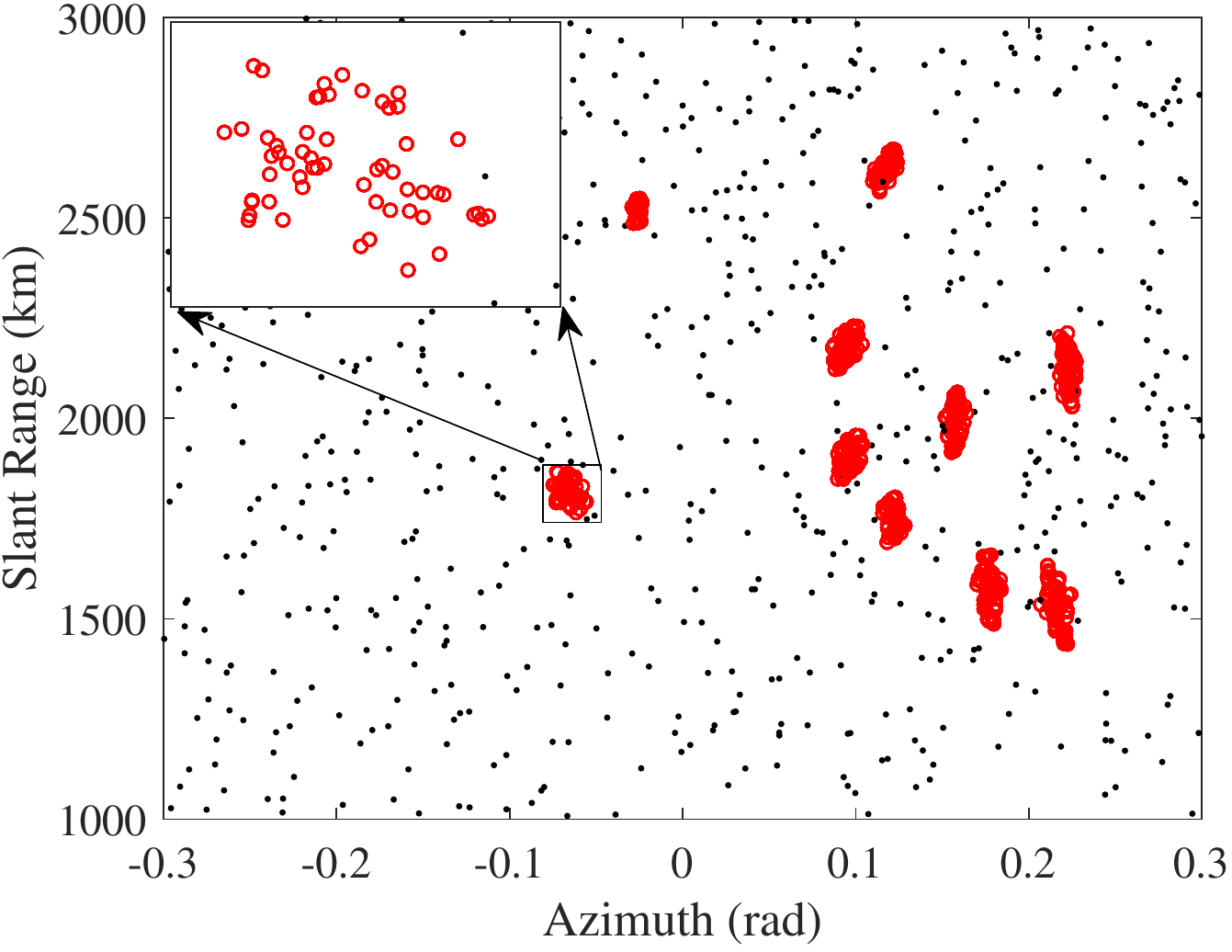}}
    \caption{Multipath measurements and clutter in slant coordinate system with $p_d = 0.4$ and $\lambda = 1e-5$. $\circ$~(red) represents multipath measurements of targets and $\cdot$~(black) represents clutter.}\label{measurement}
\end{figure}

\begin{figure}[!htbp]
\centering
    \subfloat[MP-OTHRs]{\label{fig12-a}\includegraphics[scale=0.55]{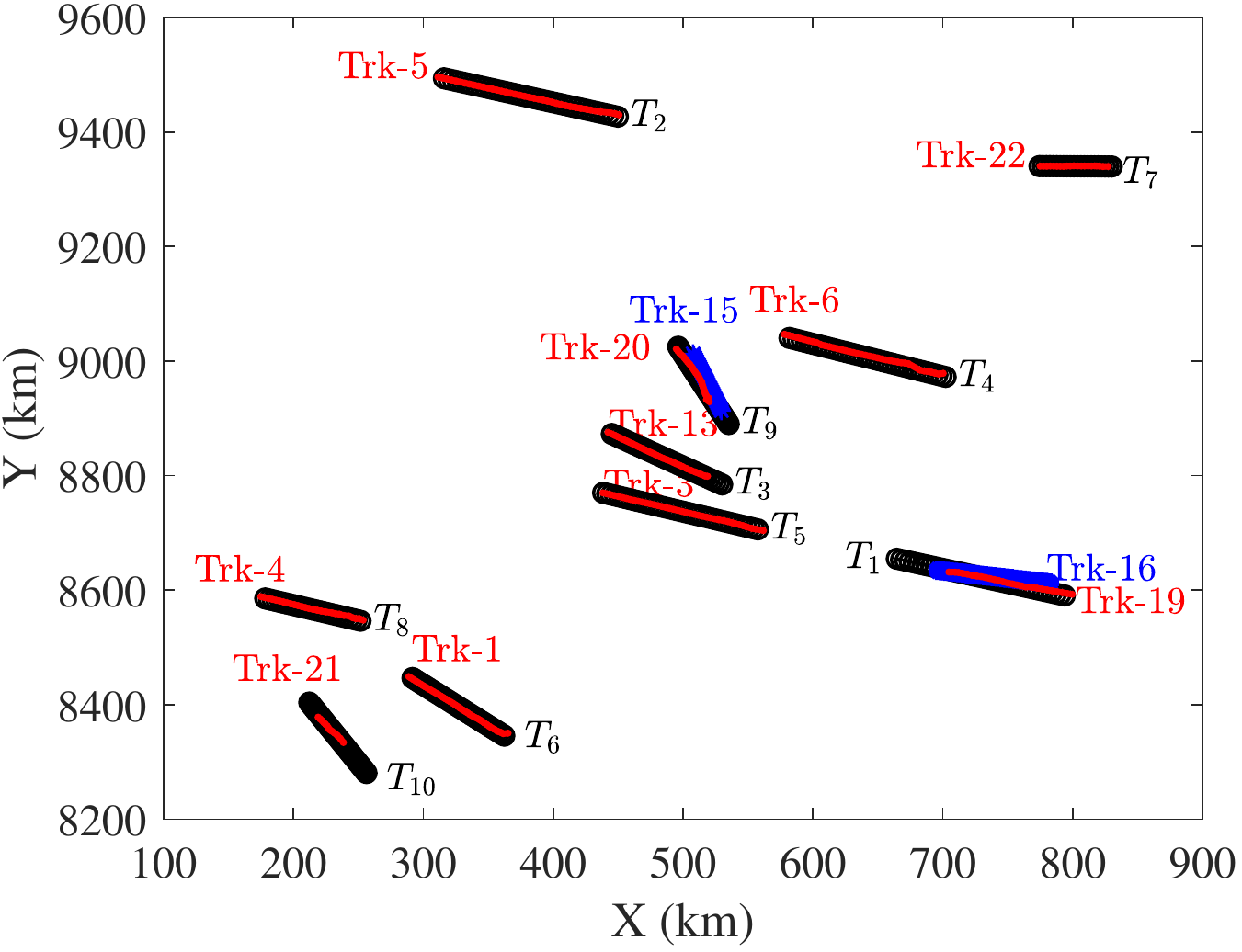}} \hspace{10pt}
    \subfloat[MP-OTHR1]{\label{fig12-b}\includegraphics[scale=0.55]{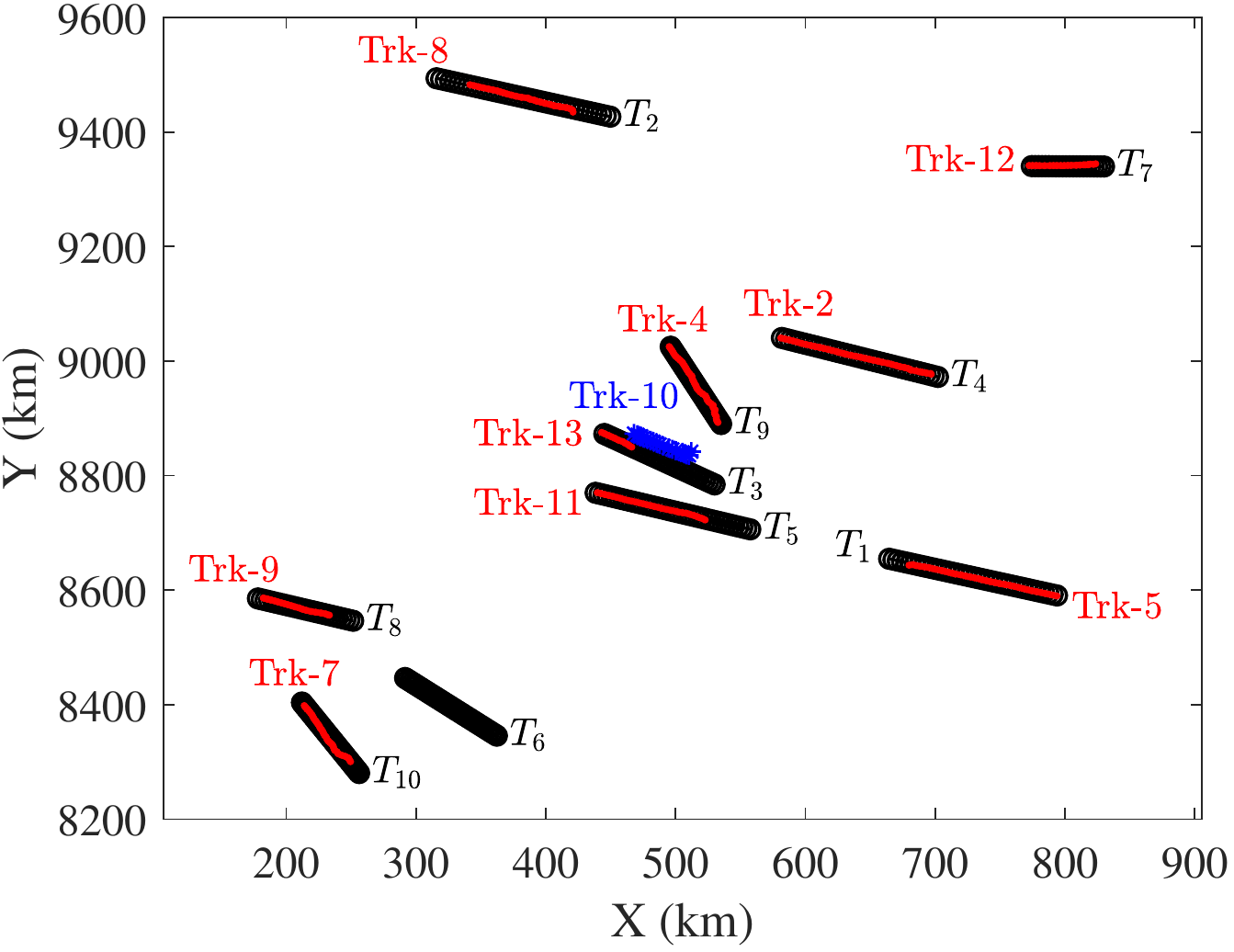}} \\
    \subfloat[MP-OTHR2]{\label{fig12-c}\includegraphics[scale=0.55]{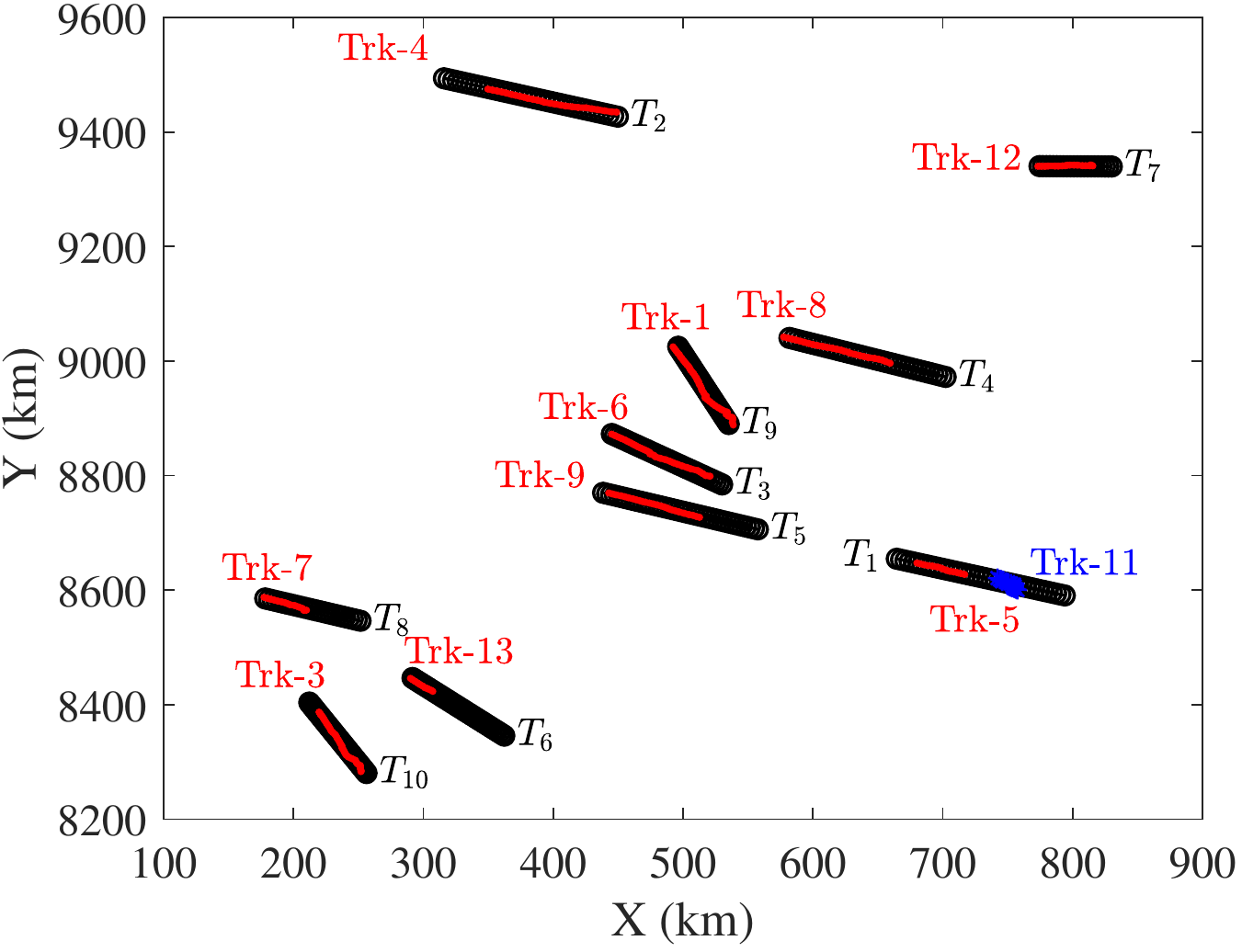}} \hspace{10pt}
    \subfloat[MR-MPTF]{\label{fig12-d}\includegraphics[scale=0.55]{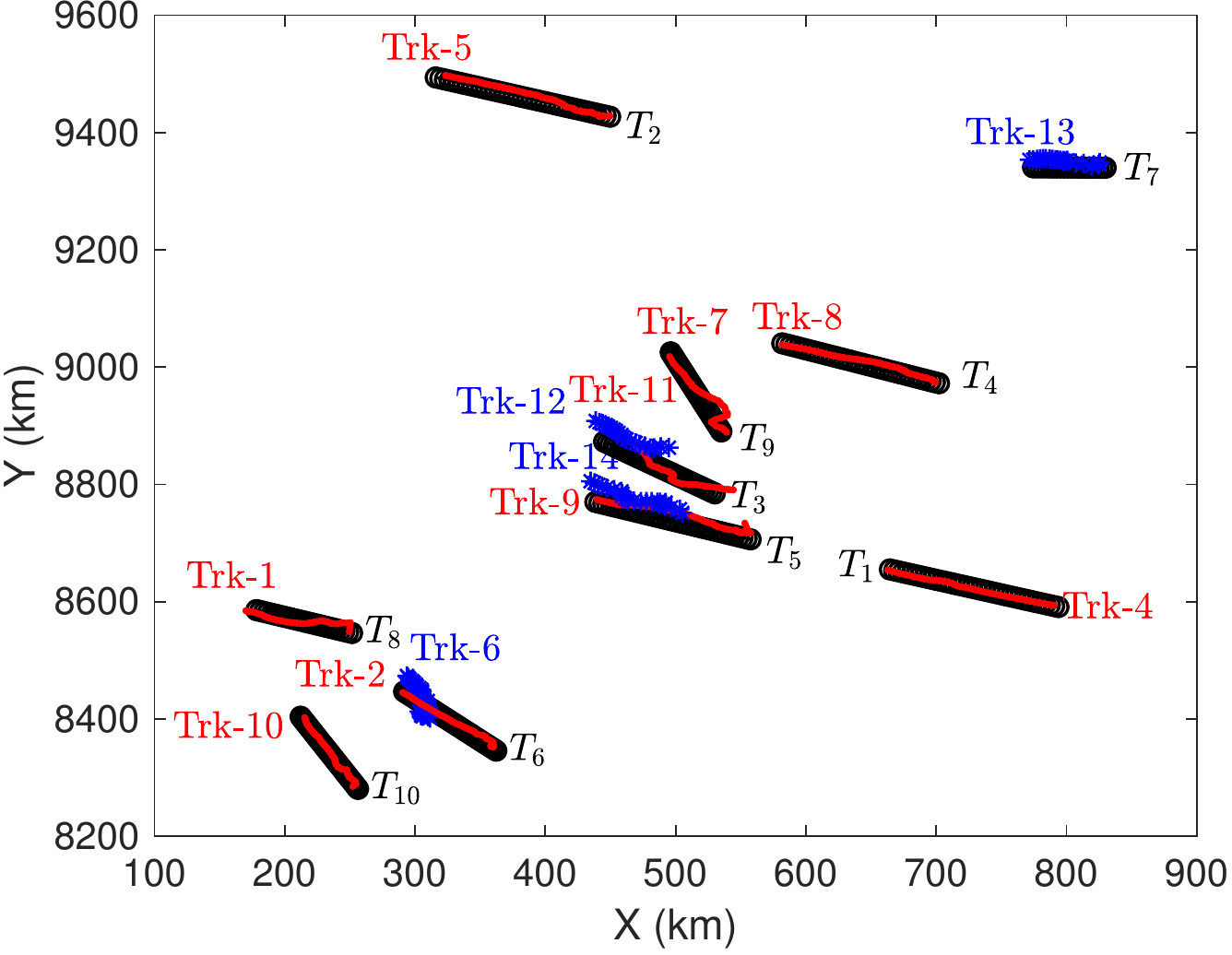}}
\caption {Tracks obtained by MP-OTHRs, MP-OTHR1, MP-OTHR2, and MR-MPTF. $\circ$~(black), $\cdot$~(red), and $\ast$~(blue) represent true trajectories, valid tracks, and false tracks, respectively.}
\label{Tracking}
\end{figure}

\subsection{Simulation Results}
Fig.~\ref{measurement} shows the multipath detections of ten targets and clutter over all scans when $P_d^{\tau, s} = 0.4$
and $\lambda^{s} = 1e-5$, $\tau = 1, \ldots, 4$, $s = 1, 2$.
The trajectories obtained by MP-OTHRs, MP-OTHR1, MP-OTHR2 and MR-MPTF in a single run are shown in Fig.~\ref{Tracking}.
From Fig.~\ref{fig12-a}, it is seen that MP-OTHRs successfully tracks all ten targets~($T_1$ - $T_{10}$) but with two false tracks~(Trk-15 and Trk-16).
Actually, the false tracks Trk-15 and Trk-16 are the multipath tracks of targets $T_9$ and $T_1$, respectively.
The tracking results of MP-OTHRs with a single OTHR are shown in Figs.~\ref{fig12-b}-~\ref{fig12-c}.
It is seen that MP-OTHR1 successfully tracks nine targets~($T_1$ - $T_5$ and $T_7$ - $T_{10}$) with one target~($T_6$) being missed, and one false track~(Trk-10, the multipath tracks of target $T_3$).
MP-OTHR2 successfully tracks all ten targets with one false track~(Trk-11, the multipath track of target $T_1$).
Comparing Fig.~\ref{fig12-a} with the Figs.~\ref{fig12-b}-~\ref{fig12-c}, the overall performance of MP-OTHRs is superior to MP-OTHR1 and MP-OTHR2. Fig.~\ref{fig12-d} shows that MR-MPTF successfully tracks nine targets~($T_1$ - $T_6$ and $T_8$ - $T_{10}$) with one target $T_7$ being missed, and four (multipath) false tracks~(Trk-6, Trk-12, Trk-13, Trk-14).
MP-OTHRs is superior to MR-MPTF in the aspects of both target detection and tracking.
All the algorithms are implemented in MATLAB R2018a on a PC with an Intel Core i5CPU and 8GB RAM.
TET of MP-OTHRs, MP-OTHR1, MP-OTHR2 and MR-MPTF are 14.6~s, 4.21~s, 3.76~s and 77.6~s, respectively.

\begin{figure}
    \centering
    \subfloat[NVT]{\label{fig13-a}\includegraphics[scale=0.45]{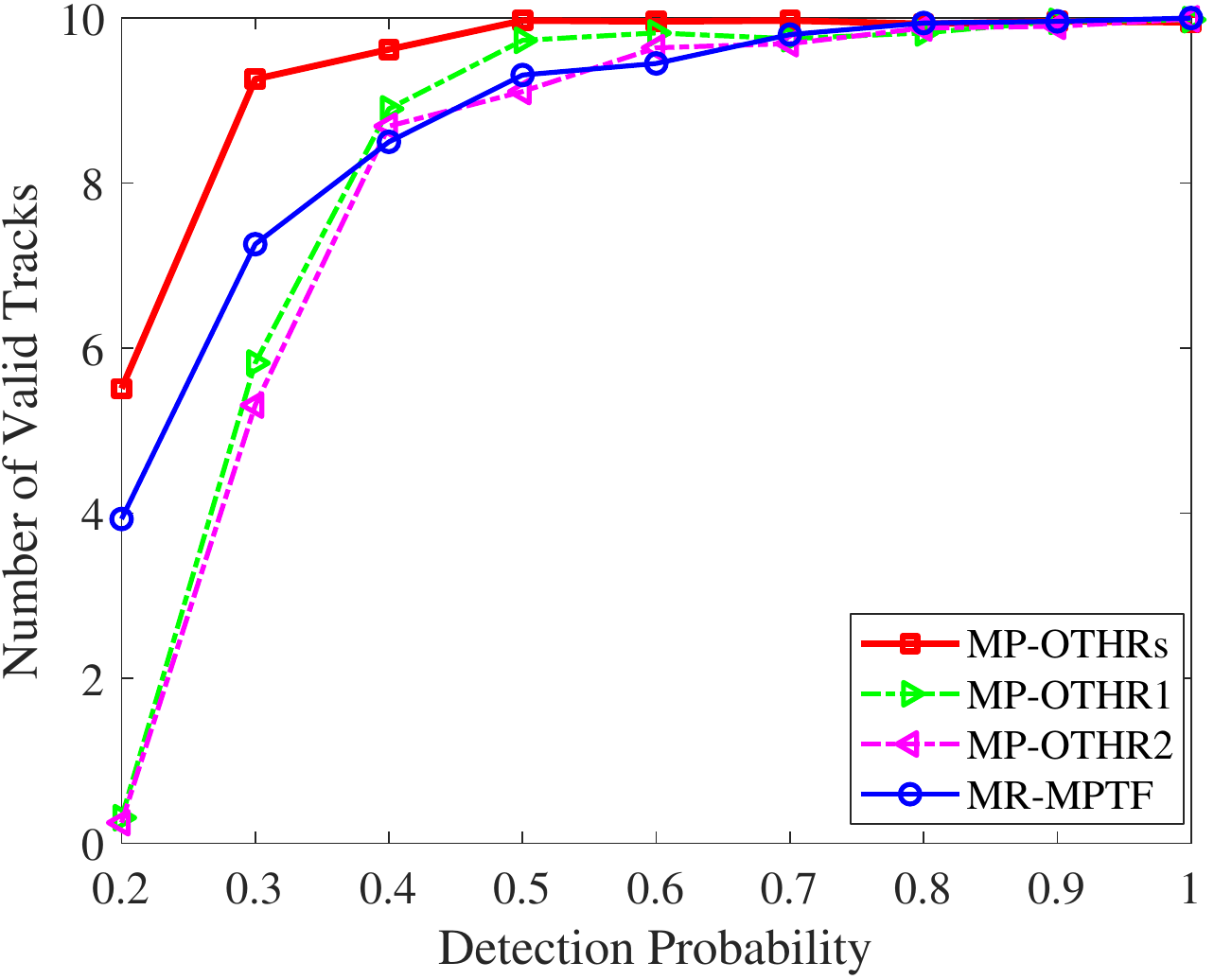}} \hspace{25pt}
    \subfloat[NFT]{\label{fig13-b}\includegraphics[scale=0.45]{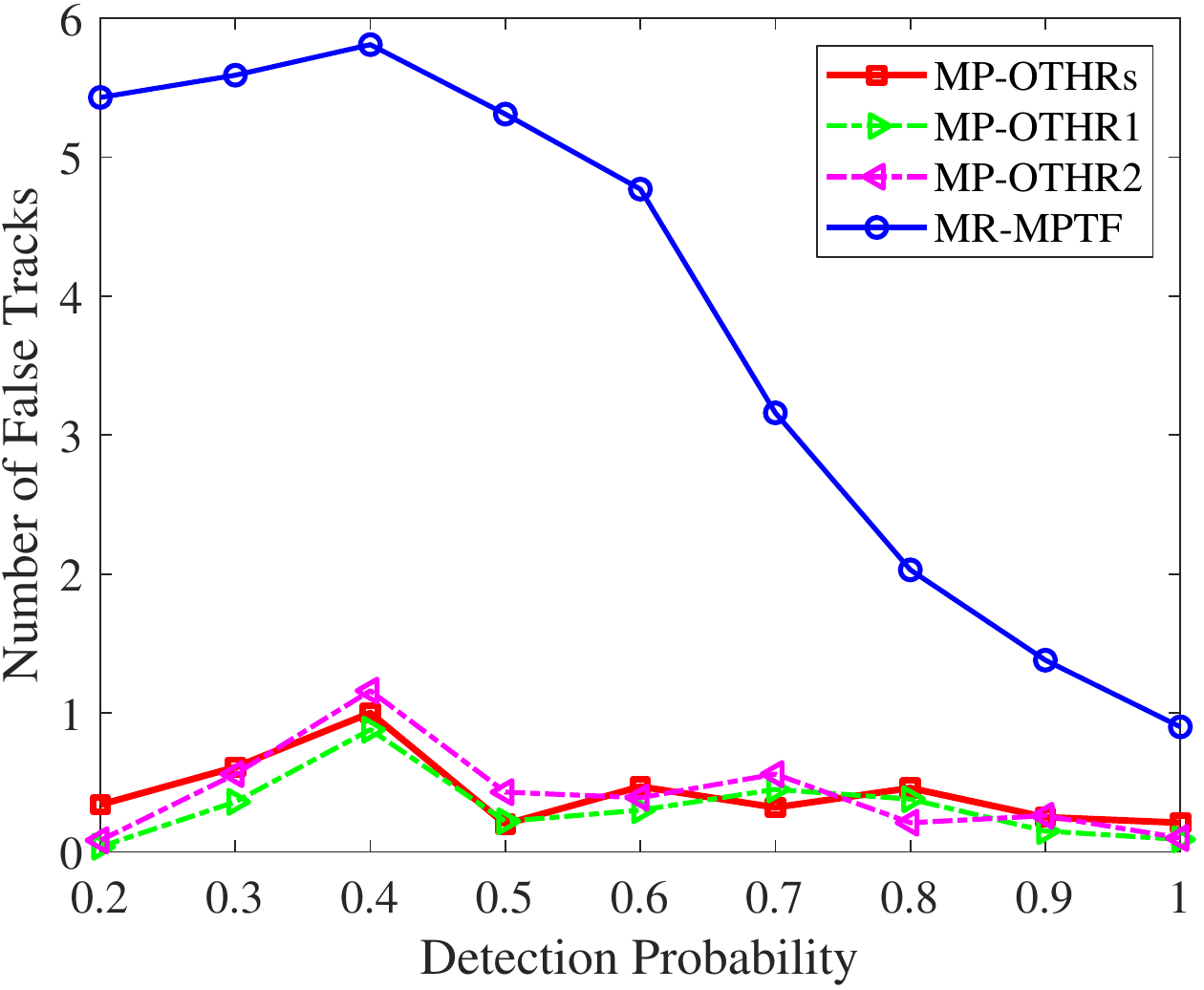}} \\
    \subfloat[TPD]{\label{fig13-c}\includegraphics[scale=0.45]{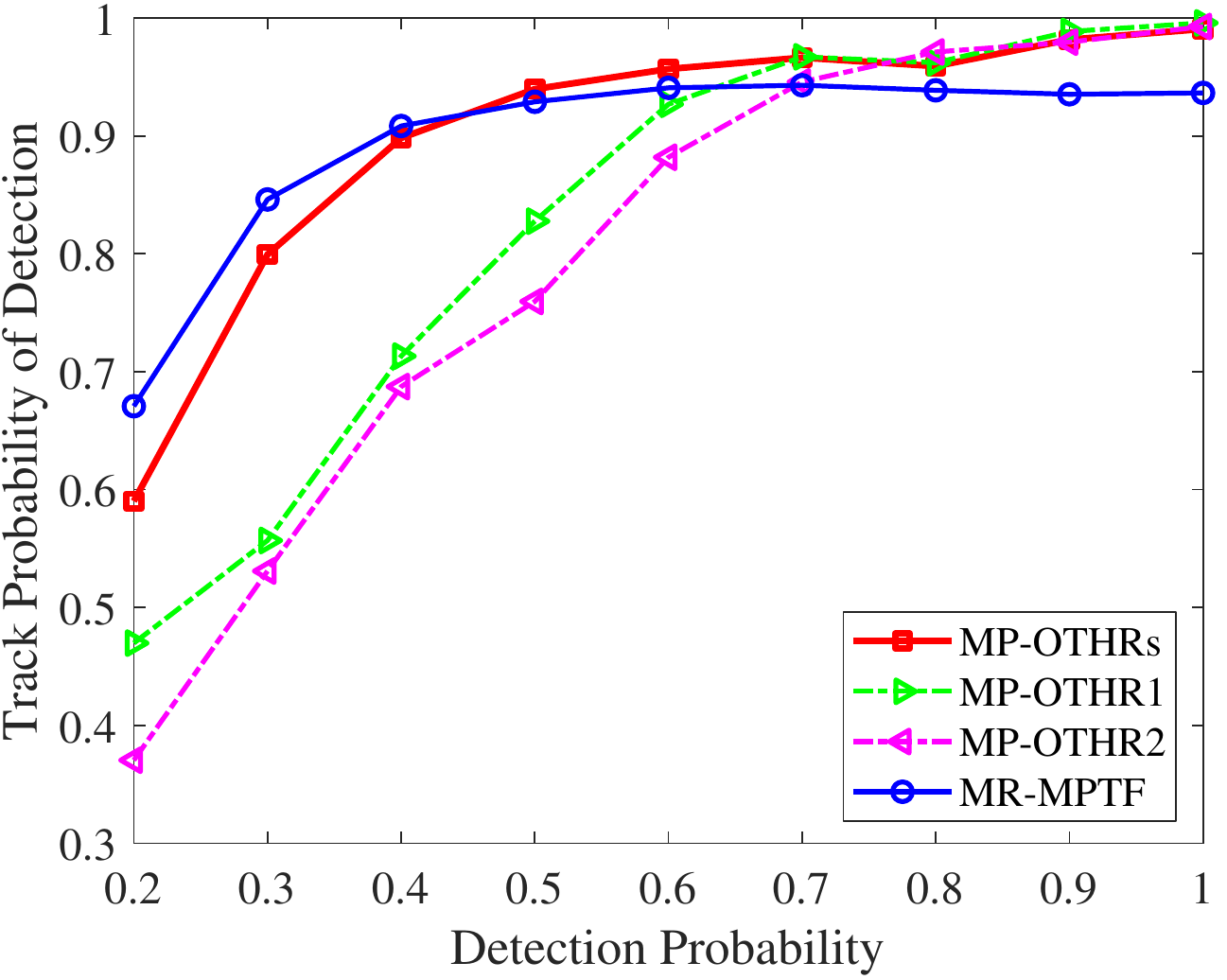}} \hspace{25pt}
    \subfloat[CLT]{\label{fig13-d}\includegraphics[scale=0.45]{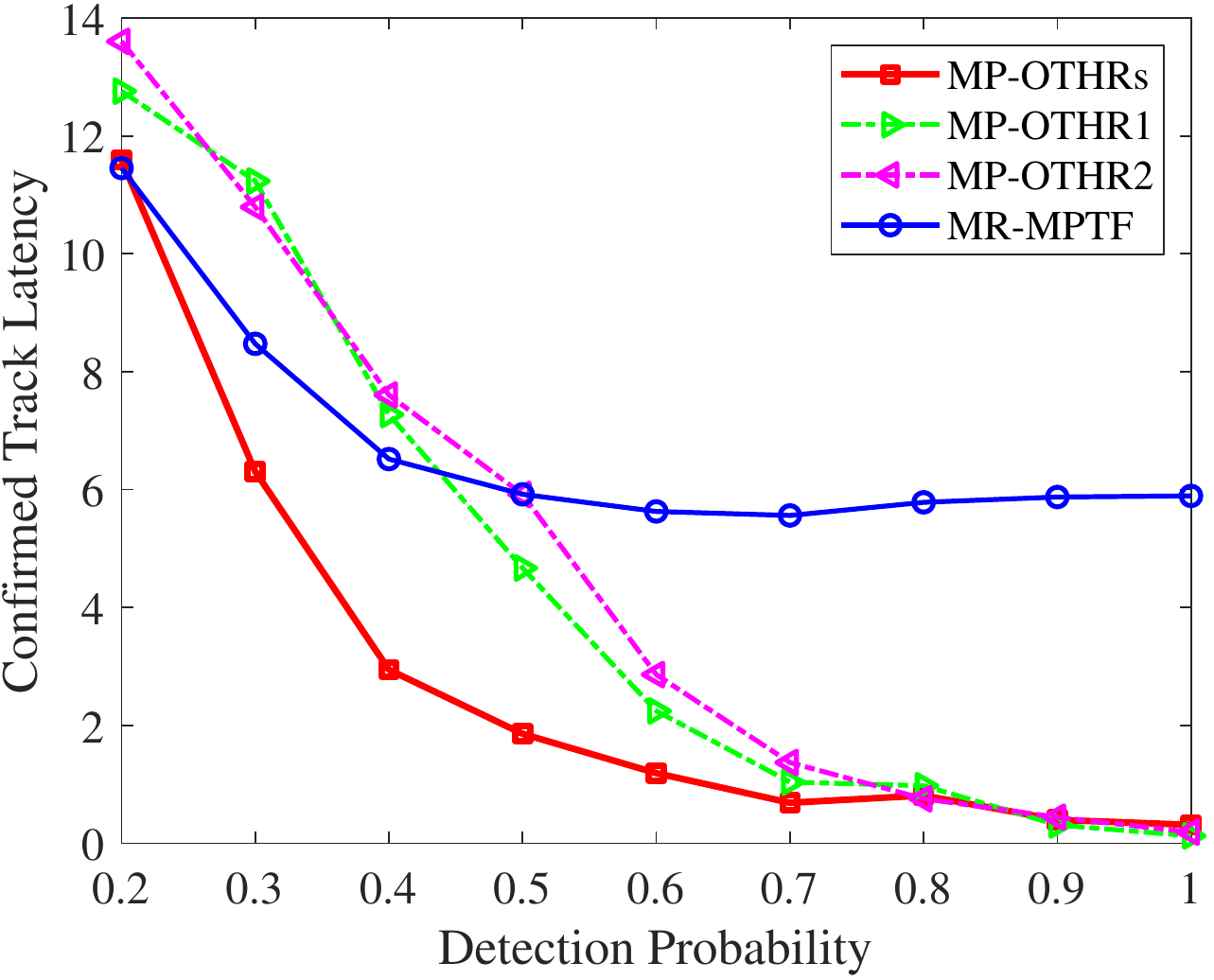}} \\
    \subfloat[AEEP]{\label{fig13-e}\includegraphics[scale=0.45]{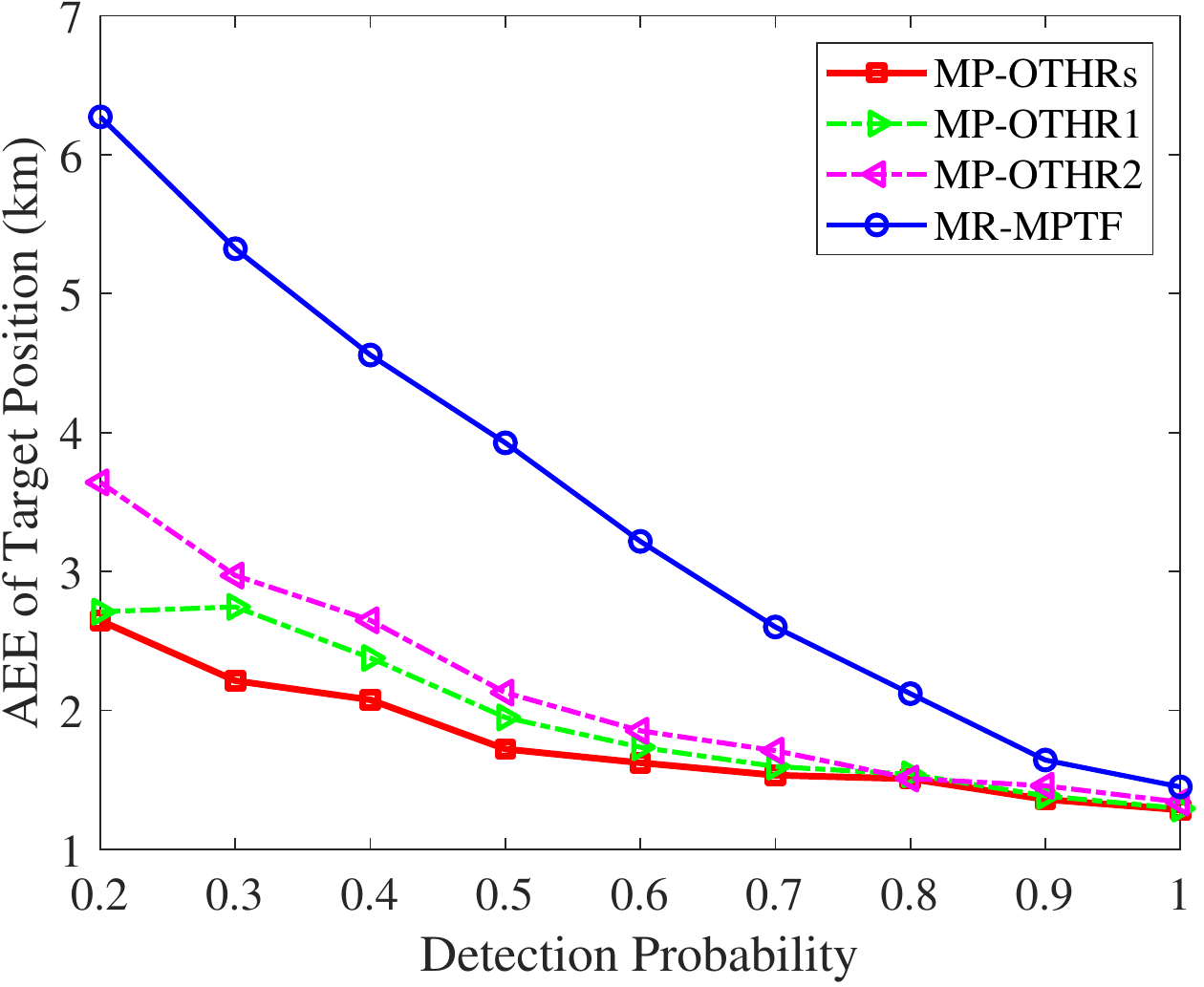}} \hspace{25pt}
    \subfloat[AEES]{\label{fig13-f}\includegraphics[scale=0.45]{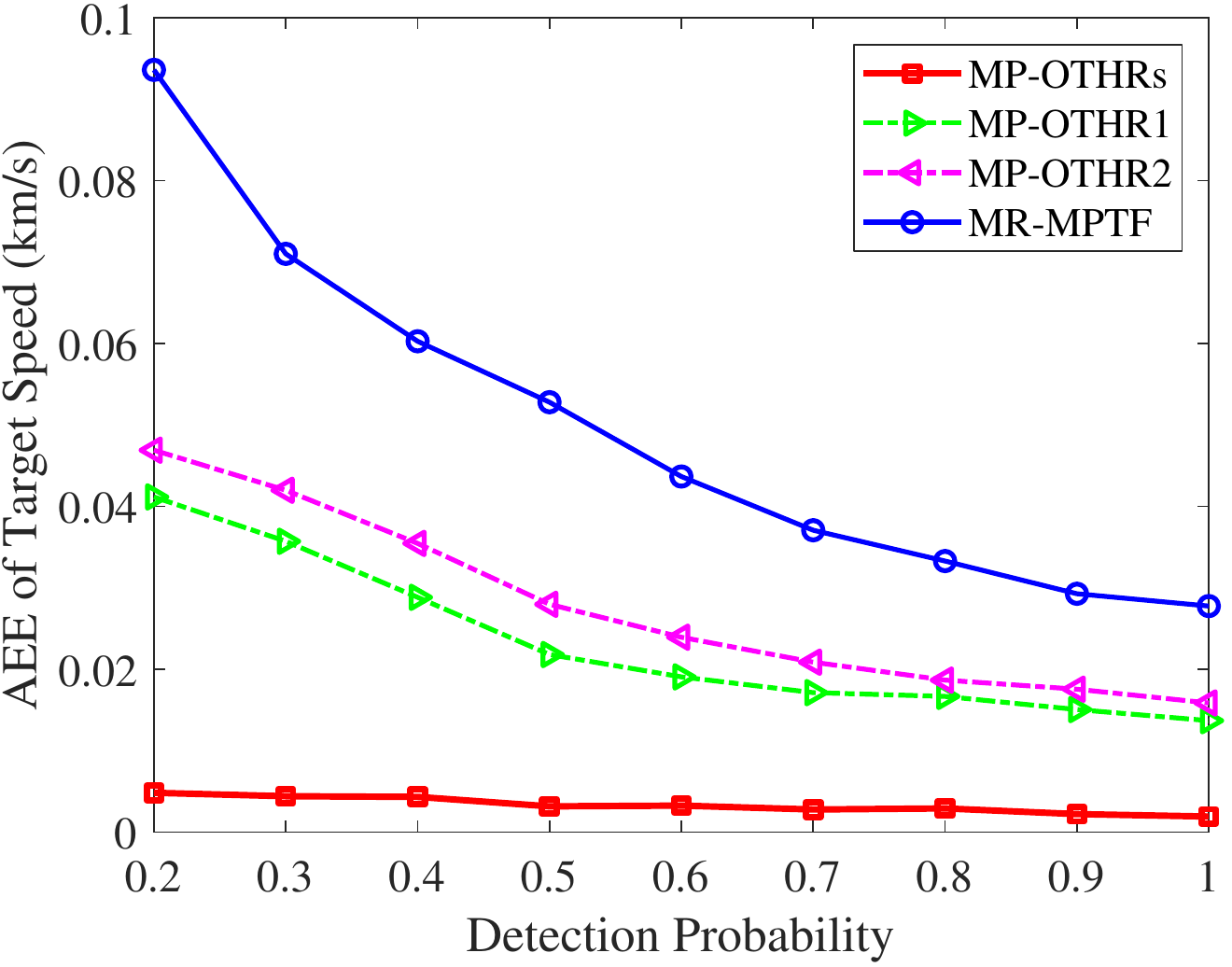}} \\
    \subfloat[AEEH]{\label{fig13-g}\includegraphics[scale=0.45]{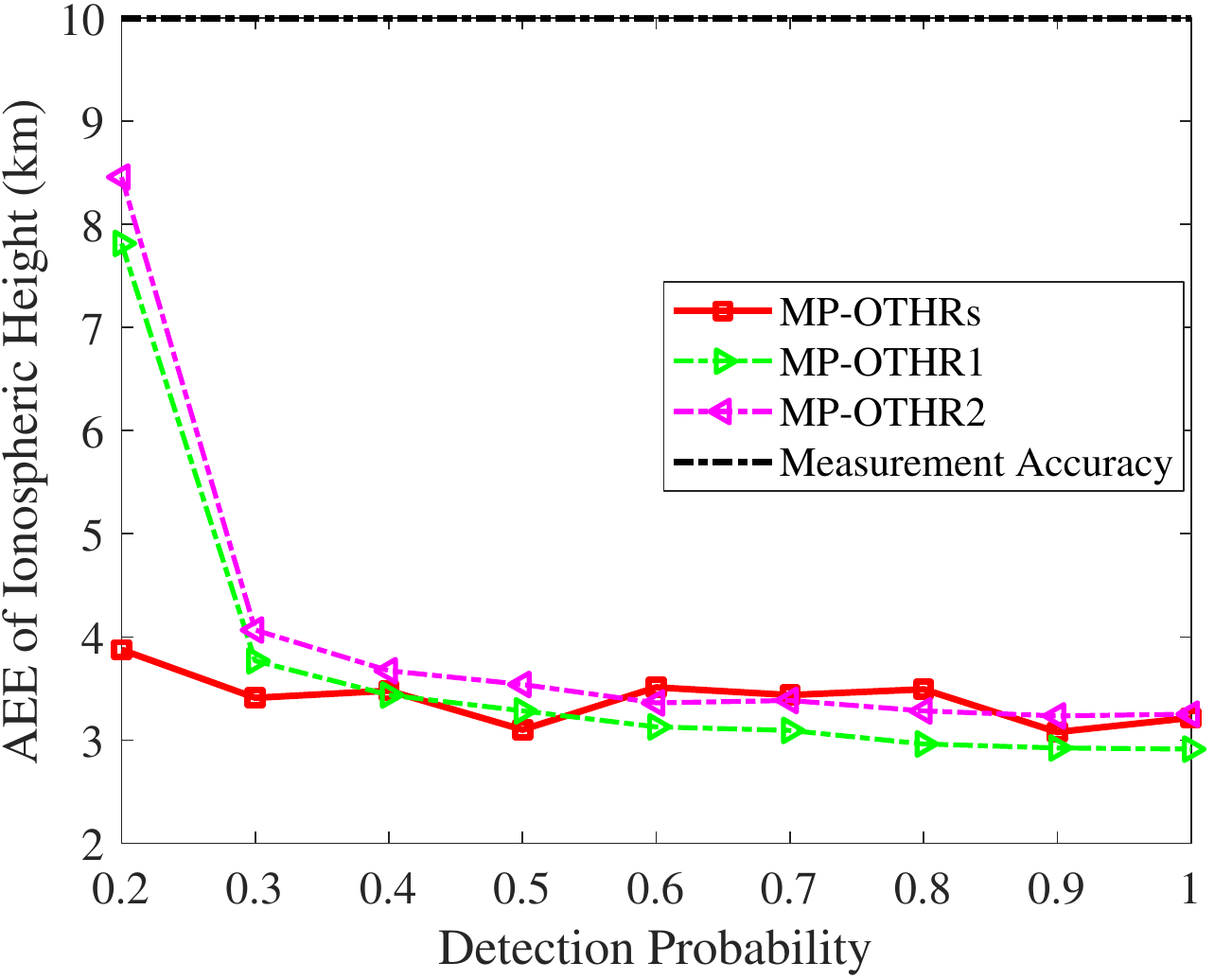}} \hspace{25pt}
    \subfloat[MOSPA]{\label{fig13-h}\includegraphics[scale=0.45]{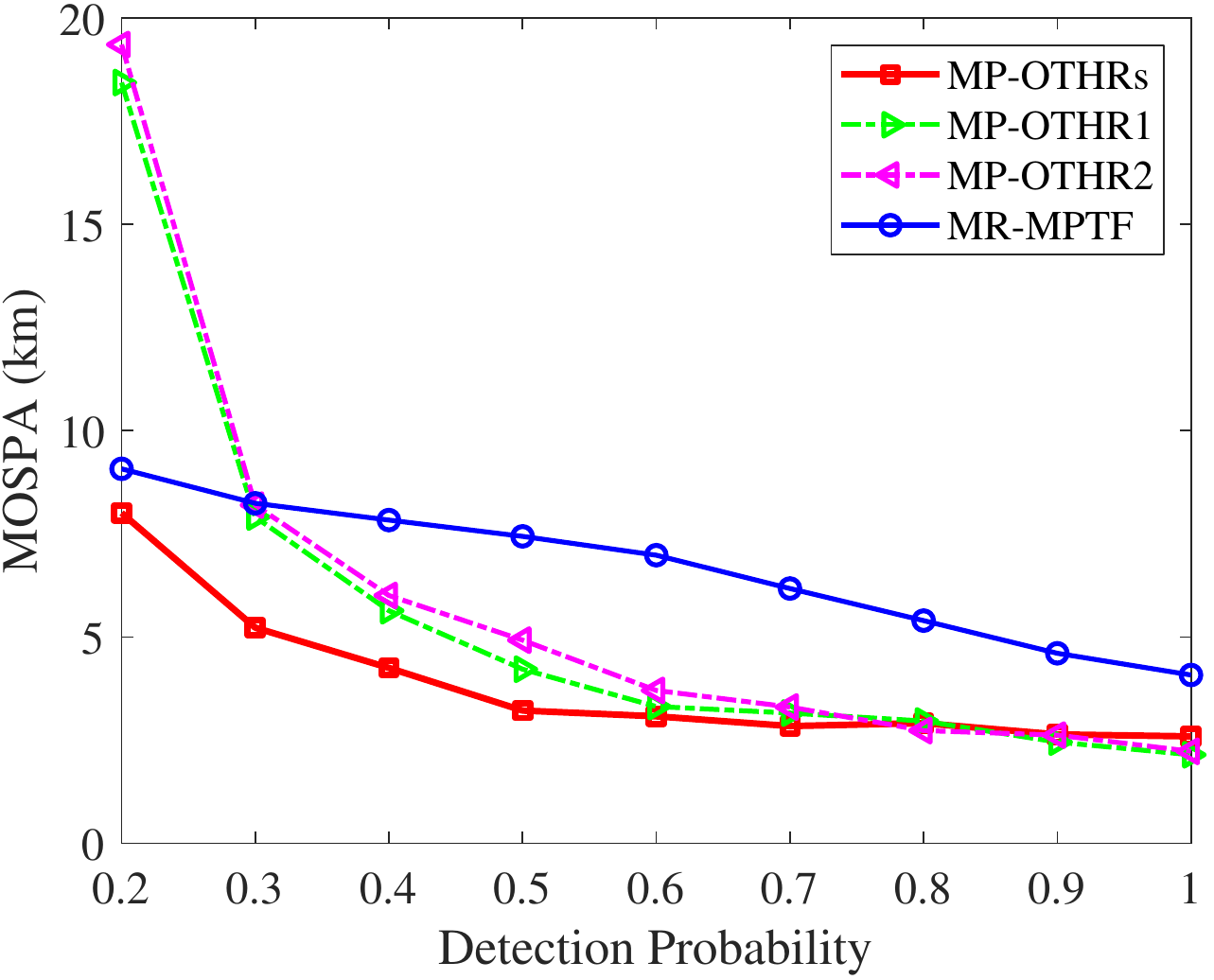}}
    \caption{Performance comparison w.~r.~t. different detection probabilities.}\label{JORN}
\end{figure}

Fig.~\ref{PCDP} shows the performance comparison w.~r.~t. different detection probability~$p_d^{\tau,s}$ with 100 Monte Carlo runs.
As expected, the performance of both target detection and tracking is improved as the increase of detection probability.
Specifically, in terms of NVT~(shown in Fig.~\ref{fig13-a}), under the extremely low detection probability $p_d^{\tau,s} = 0.2$, it is hard for a single OTHR, i.e., MP-OTHR1 or MP-OTHR2, to detect valid tracks.
Integrating information from the OTHR network, MP-OTHRs and MR-MPTF can track about half of the total targets,
and the NVT of MP-OTHRs is greater than that of MR-MPTF.
This is because MR-MPTF is based on track-level fusion whereas multipath track is hard to be initialized in slant coordinate system and/or multipath track fusion is unreliable under low detection probability circumstance.
MP-OTHRs is based on measurement-level fusion which is more beneficial to track maintenance.
As the detection probability increased~(e.g., $p_d^{\tau,s} \geq 0.5$), NVTs of MP-OTHRs, MP-OTHR1, MP-OTHR2 and MR-MPTF are becoming comparable. This is because the multipath tracks are easy to be detected by MR-MPTF if $p_d^{\tau,s}$ is high.
For a single OTHR, if $p_d^{\tau,s} = 0.5$, the probability of at least one (path) detection for a target is 0.875; the NVTs of MP-OTHR1 and MP-OTHR2 can be improved greatly.
In the aspect of NFT~(shown in Fig.~\ref{fig13-b}), which is mainly originated from the multipath tracks, MP-OTHRs performs better than MR-MPTF, especially in the low detection probability cases.
With the same reason on NVT, MR-MPTF is easily to produce false tracks.
By adopting global track initialization as described in Section \ref{sec:initialization}, MP-OTHRs does not generate many false tracks.
In the aspects of TPD~(shown in Fig.~\ref{fig13-c}) and CLT~(shown in Fig.~\ref{fig13-d}),
MP-OTHRs outperforms MP-OTHR1 and MP-OTHR2 in the case of low detection probability since
using all the measurements from the OTHR network is benefit to stable target tracking and fast track initialization.
TPD of MP-OTHRs and MR-MPTF are comparable, and MP-OTHRs is superior to MR-MPTF on CLT.
This is because MR-MPTF adopts the 3/5 logic to confirm tracks while MP-OTHRs uses visibility probability to fast track confirmation.
On the tracking error~(shown in Figs.~\ref{fig13-e}-\ref{fig13-f}), MP-OTHRs has the best tracking accuracy, while MR-MPTF is worst.
The reason is that MP-OTHRs adopts the iteration mechanism and state smooth using a batch of measurements;
MR-MPTF adopts a filter without iteration and state smooth. Meanwhile, MP-OTHRs is superior to MP-OTHR1 and MP-OTHR2 because of using the measurements from independent OTHRs.
As shown in Fig.~\ref{fig13-g}, the error on ionospheric height is reduced compared with the ionospheric measurement error.
This improvement is achieved by the information exchange between local ionospheric height identification and global target state estimation.
The improved target track accuracy aids in the identification of ionospheric height.
The MSOPA curves~(Fig.~\ref{fig13-h}) show that, on the whole, MP-OTHRs is superior to MR-MPTF.
Meanwhile, MP-OTHRs is superior to MP-OTHR1 and MP-OTHR2 in the low detection probability cases.

 \label{PCDP}

\section{Conclusion}\label{sec:conclusion}
We studied target tracking and fusion for an OTHR network.
Based on MP, we proposed a joint optimization algorithm for OTHR measurement-level fusion, MP-OTHRs, which is a closed-loop solution among target detection, target tracking, multipath data association and ionospheric height identification.
MP-OTHRs improves the performance of target detection and tracking significantly comparing with the track-level fusion method, MR-MPTF.
Compared with a single OTHR, MP-OTHRs improves target detection and tracking performance, especially in the low detection probability cases.

\section{Acknowledgments}
This work was supported in part by the National Natural Science Foundation of China (Grant No.~61873211, 61501378, 61503305, 61790552).

\bibliographystyle{IEEEtran}
\bibliography{IEEEabrv,VB-OTHR-Fusion} 

\end{document}